\documentclass[%
 aip,jcp,reprint,
]{revtex4-1}

\usepackage{graphicx}
\usepackage{dcolumn}
\usepackage{bm}
\usepackage{hyperref}
\usepackage{array} 
\usepackage[export]{adjustbox}
\usepackage[nomessages]{fp}
\usepackage{xcolor}
\usepackage{placeins}
\usepackage[version=3]{mhchem} 
\usepackage[T1]{fontenc}       

\begin{document}

\preprint{AIP/123-QED}

\author{Kun Yao}
\author{John E. Herr}
\author{David W. Toth}
\author{Ryker Mcintyre}
\author{John Parkhill}%
\email{john.parkhill@gmail.com}
\affiliation{%
 Dept. of Chemistry and Biochemistry, The University of Notre Dame du Lac 
}%

\date{\today}

\title{The TensorMol-0.1 Model Chemistry: a Neural Network Augmented with Long-Range Physics.}

\begin{abstract}
Traditional force-fields cannot model chemical reactivity, and suffer from low generality without re-fitting. Neural network potentials promise to address these problems, offering energies and forces with near ab-initio accuracy at low cost. However a data-driven approach is naturally inefficient for long-range interatomic forces that have simple physical formulas. In this manuscript we construct a hybrid model chemistry consisting of a nearsighted Neural-Network potential with screened long-range electrostatic and Van-Der-Waals physics. This trained potential, simply dubbed "TensorMol-0.1", is offered in an open-source python package capable of many of the simulation types commonly used to study chemistry: Geometry optimizations, harmonic spectra, and open or periodic molecular dynamics, Monte Carlo, and nudged elastic band calculations. We describe the robustness and speed of the package, demonstrating millihartree accuracy and scalability to tens-of-thousands of atoms on ordinary laptops. We demonstrate the performance of the model by reproducing vibrational spectra, and simulating molecular dynamics of a protein. Our comparisons with electronic structure theory and experiment demonstrate that neural network molecular dynamics is poised to become an important tool for molecular simulation, lowering the resource barrier to simulate chemistry. 

\end{abstract}

\maketitle

\section{\label{sec:level1}Introduction}
\indent Neural network model chemistries (NNMCs) greatly reduce the computational effort needed to simulate chemical systems with ab-initio accuracy\cite{Snyder:2013aa,brockherde2017bypassing,snyder2012finding,li2016understanding,li2016pure,vu2015understanding,handley2010potential,chmiela2017machine,behler2011neural,behler2007generalized,shakouri2017accurate,behler2017first, han2017deep,yao2017many,yao2016kinetic,yao2017intrinsic,khaliullin2011nucleation,bartok2010gaussian, mones2016exploration,gastegger2017machine,kobayashi2017neural, carpenter2017empirical,kolb2017discovering,kruglov2017energy, lubbers2017hierarchical,mills2017deep,wu2017internal,khorshidi2016amp,shao2016communication,zhang2014effects, li2015permutationally2,medders2015representation,medders2013critical,reddy2016accuracy,riera2017toward,moberg2017molecular, conte2015permutationally,manzhos2014neural, manzhos2009fitting,malshe2010input,peterson2016acceleration, khorshidi2016amp,piquemal2017preface,cubuk2017representations, MBCoarseCsanyi,FFparafit,li2017machine}. They can be used to predict molecular properties \cite{CM, BoBs,lopez2014modeling, pilania2013accelerating, schutt2014represent, ma2015machine, nelson2012nonadiabatic, janet2017predicting, janet2017resolving, yao2017intrinsic, hase2017machine, mcgibbon2017improving, bereau2017non, grisafi2017symmetry, isayev2017universal, ghiringhelli2017learning, ouyang2017sisso, faber2017prediction, schutt2017quantum, MSLSTM, timoshenko2017supervised}, and design new materials or drugs\cite{li2017learning, ramsundar2017multitask, hachmann2011harvard,hachmann2014lead, isayev2015materials, kim2017machine, segler2017towards,olivares2011accelerated,guimaraes2017objective,wei2016neural, gomez2016automatic,jinnouchi2017predicting,ulissi2017machine,sun2017machine}. In spite of growing popularity, most neural network methods are still only used by their developers and are customized for a single application. The paucity of literature describing transferable accuracy, confusion about what physics can be reproduced, and dearth of open software is slowing adoption. This paper develops an open-source, transferable neural network model chemistry. We show that NNMCs are easily hybridized with physical contributions to molecular energies familiar from Molecular Mechanics and corrections to Density Functional Theory(DFT)\cite{grimme2006semiempirical}. This approach combines the best of both worlds, yielding predictable reproduction of physical long-range forces, but also featuring a linear-scaling inductive charge model which is cheaper than a Thole model\cite{thole1981molecular} yet more accurate than fixed charges. \\ 
\indent     Our group is one of several who have been pursuing transferable and black-box neural network model chemistries\cite{yao2017intrinsic, brockherde2017bypassing, smith2017ani, medders2013critical, behler2007generalized, han2017deep}. The field is growing so rapidly, that most non-practitioners cannot keep up with the capabilities of existing models and the outstanding problems. Readers may not appreciate that a model can achieve chemical accuracy for energies but have uselessly noisy forces. Models which provide energies at equilibrium, and those treating a fixed molecule or stoichiometry are now reliably produced\cite{han2017deep}. We will show that TensorMol-0.1 yields usefully accurate predictions of forces out-of-equilibrium by showing reproduction of infrared spectra which closely approximate our source model chemistry (wB97X-D, 6-311G**)\cite{chai2008long}, and molecular dynamics. We outline several tricks which are required to ensure the stability of long-time molecular dynamics.\\
\indent     Another distinguishing feature of our approach is the open-source nature of our package, which uses the TensorFlow tensor algebra system to compute descriptors and forces. Rather than a monolithic black-box TensorMol-0.1 is a modular collection of differentiable chemical models written in concise Python code. The components are easily joined together and extended. The methodology can be used to propagate dynamics for large molecules ($10^5$ atoms) with GPU acceleration on simple laptop computers. No significant expertise, force field refinement, or other interventions are needed to apply the method to a molecule of interest, so long as the elements are supported. The package is also interfaced with the I-PI path integral package\cite{ceriotti2014pi}, to allow for quantum simulations and enhanced sampling. \\

\section{\label{sec:level1}Methods}

\begin{figure}[t]
\includegraphics[width=0.45\textwidth]{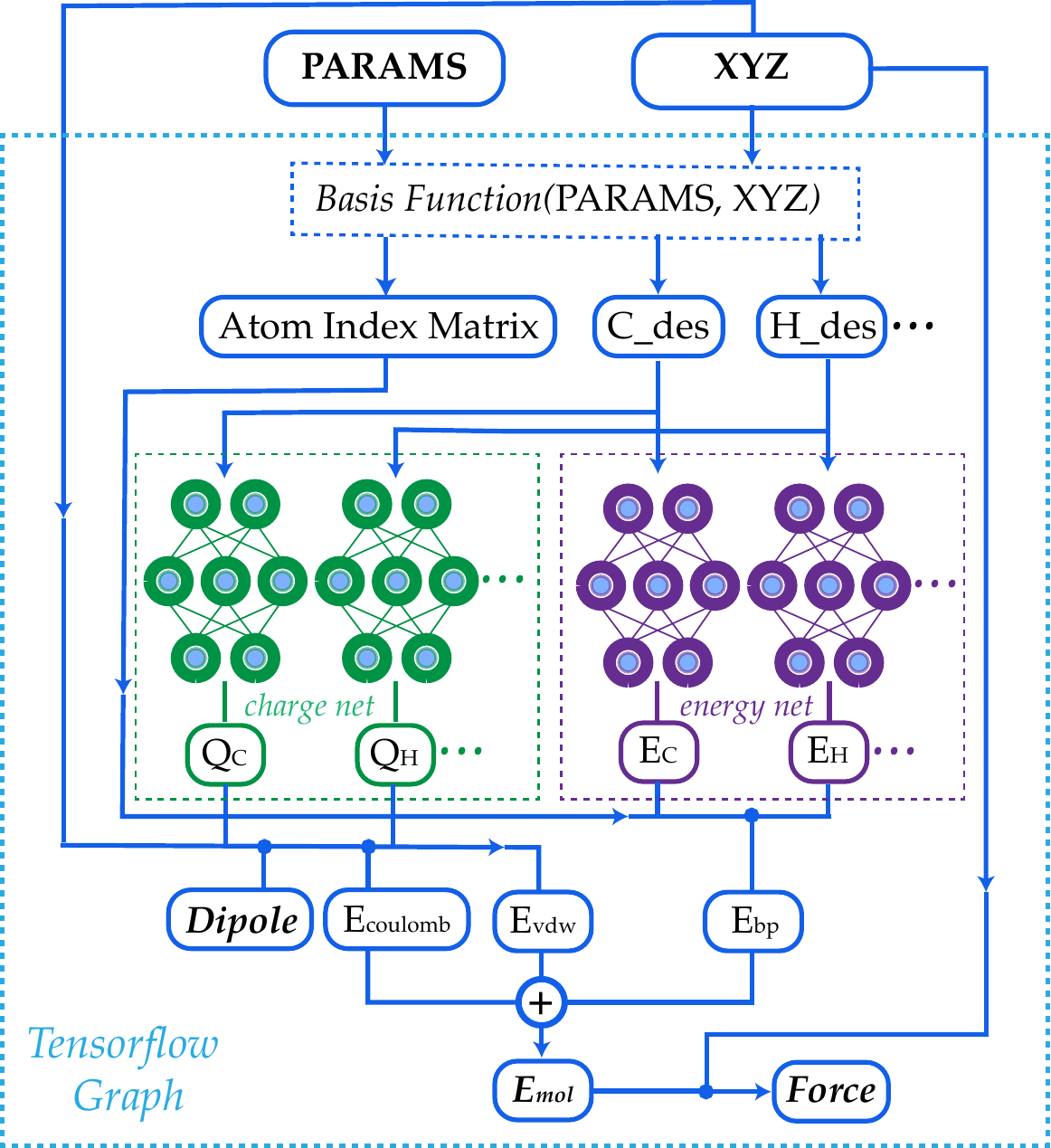}
\caption{The schematic graph of TensorMol-0.1. Each element has its own charge network and energy network. The charge network predicts atomic charges that yield the ab-initio dipole moment. The Behler-Parinello type energy network produces a short-range embedded atomic energy, which is summed with the electrostatic energy and Van-Der Waals energy to predict the total atomization energy of molecules at and away from equilibrium. The whole computation is included in the TensorFlow graph. Analytical GPU-accelerated forces are symbolically generated by a single line of code.}
\label{Fig1}
\end{figure}

\indent     The community of Neural-Network model chemistry developers is rapidly improving the accuracy and generality of these reactive force fields\cite{behler2017first,  shakouri2017accurate, bartok2010gaussian, deringer2017machine, li2015permutationally2, smith2017ani}. The model of this paper includes several components which were the subject of recent developments in other groups\cite{behler2007generalized,behler2017first,morawietz2013density, smith2017ani, gastegger2017machine}. We will describe the details here from the bottom up citing prior art. Our notational convention will be that $i,j,k...$ are indices of atoms, $q_i$ is the charge on atom i, $z,x,y$ are atomic numbers, $\mathcal{A},\mathcal{B}, \mathcal{C}$ are molecules, and $\alpha, \beta... $ are indices of basis functions which are a product of radial and angular functions. If a function depends on all the atomic coordinates of a molecule it will be written as a vector, and those which depend on only few will be given explicit indices. The energy of TensorMol-0.1 is expressed as a sum of a short-range embedded N-body potential\cite{behler2017first}, and long-range electrostatic potential and Van-Der-Waals force: 
\begin{eqnarray}
E(\vec{R}) =&&\sum_i E^\text{BP}_{z_i} (S_\alpha(\vec{R})) +\sum_{ij} E^\text{DSF}_{ij}(q_{ij}(S_\alpha(\vec{R})),R_i,R_j)\nonumber\\&&+E^\text{VDW}(\vec{R}_{ij})
\end{eqnarray}
In the above expression $E_{z_i}$ is a Behler-Parinello type energy network for the element $z$ for atom $i$. This n-body potential takes as its argument $S_\alpha$ the modified symmetry functions of Isayev and coworkers\cite{smith2017ani}: 
\begin{eqnarray}
S_\alpha (radial) = && \sum_{j\neq i}e^{{-\eta(R_{ij}-R_{s})^2}}f_c(R_{ij}) \\
S_\alpha (angular) =&& 2^{1-\zeta }\sum_{j\neq i, j\neq k}(1+cos(\theta_{ijk}-\theta_s))^{\zeta}\nonumber\\&&\times e^{-\eta(\frac{R_{ij}+R_{ik}}{2}-R_s)^2}f_c(R_{ij})f_c(R_{ik})
\end{eqnarray}
Modern machine learning frameworks provide automatic differentiation of tensor algebraic expressions, allowing a force-field developer to obtain the gradient of a molecular potential $\frac{dE(\vec{R})}{d\vec{R}}$ in a single line of code, once the expression for $E(\vec{R})$ has been written. An important feature of our code is that this symmetry function is coded within the TensorFlow system\cite{tensorflow2015-whitepaper}, so all the parameters of this descriptor can be variationally optimized alongside the network weights. Our implementation of the symmetry function employs a list of nearest-pairs and triples within radial cutoffs such that the scaling of the overall network is asymptotically linear. On an ordinary laptop equipped with only a CPU a force/energy call on 20,000 atoms is less than a minute. \\
\indent     The second term of our energy expression is the Damped-Shifted Force (DSF) Coulomb energy of Gezelter and coworkers\cite{fennell2006ewald}. The charges are obtained from a sub-network which reproduces molecular dipole moments. Our charge model enforces conservation of total charge by evenly spreading any required neutralizing charge over the entire molecule or unit cell. The Damped-shifted force ensures long range continuity and differentiability of the effective Coulomb potential with smooth cutoffs. We modify the DSF kernel at short range with an "elu" type non-linearity, such that the forces within the radius of the Behler-Parinello symmetry function smoothly approach zero avoiding singularities and interference with the Behler-Parinello many-body potential.
\begin{eqnarray}
E_{DSF}= &&\left\{\begin{matrix}  E_\text{DSF(Original)}  \hfill   R > R_\text{switch}
\\ q_iq_j(a_\text{elu}e^{R-R_\text{switch}}+\beta_\text{elu})\hfill R< R_\text{switch}
\end{matrix}\right.
\end{eqnarray}
where $E_\text{DSF(Original)}$ is the energy of DSF kernel\cite{fennell2006ewald},  $R_{switch}$ is the short range cutoff for the "elu" kernel. $\alpha_\text{elu}$ and $\beta_\text{elu}$ are chosen so that the value and the gradient of $E_\text{DSF}$ are continuous at $R_\text{switch}$.
The damped-shifted force is well-suited to combination with Neural Network models because it requires no Fourier transformation to treat periodic systems with linear scaling and maps well onto TensorFlow. The last term is the van-der waals energy, which is calculated by following Grimme's C6 scheme \cite{grimme2006semiempirical}.\\
\indent We employed a two step training approach. First, the charge networks are trained to learn the atom charges that predict the dipole moment. The loss function can be written as following:
\begin{align}
 L_{dipole} = &&\sum_\mathcal{A} (\frac{\mu^\text{DFT}_\mathcal{A} - \mu^{NN}_\mathcal{A}(q_i, q_j,...)}{N_\text{atom}})^2
\end{align}
After the charge training is converged, we train the energy network. During the energy network training, the weights in charge networks are kept frozen, but they are still evaluated to calculate the electrostatic energy that is added to construct the total energy. Our Behler-Parinello many-body potential also absorbs the shape of the transition between many-body and electrostatic regions. The learning target for the energy network includes both the DFT energy and DFT force. The loss function for the energy network training is:
\begin{align}
 L_\text{energy} =&& \sum_\mathcal{A} (\frac{E^\text{DFT}_\mathcal{A} - E^\text{NN}_\mathcal{A}}{N_{atom}})^2 + \gamma \sum_\mathcal{A} (\frac{F^\text{DFT}_\mathcal{A} - F^\text{NN}_\mathcal{A}}{N_\text{atom}})^2
\end{align}
where $E^\text{NN}$ is obtained according to equation 1, $F^\text{NN}$ is calculated by taking the gradient of $E^\text{NN}$ respect to the coordinates of the atoms. $N_\text{atom}$ is the number of the atoms in the system and $\gamma$ is a parameter that controlling the portion of force loss. We employ $\gamma=0.05$. 
\indent  We trained two neural networks based on two sets of data. One network ("water network") was trained on a database that includes $\sim$370,000 water clusters with 1 water molecule to 21 water molecules. The other network was trained on $\sim$3,000,000 different geometries of $\sim$ 15,000 different molecules that only contains C, H, O and N and up to 35 atoms. Since these 15K molecules were sampled randomly from the chemspider database, therefore we will refer this network as "chemspider network" in the following text. The training data were sampled using metadynamics and calculated by Qchem package\cite{shao2015advances} with WB97X-D\cite{chai2008long},  exchange correlation functional and 6-311G** basis set.\\

Each charge network and energy network contains three fully-connected hidden layers with 500 hidden neurons in each layer. For chemspider network, a network with three hidden layers with 2000 hidden neurons in each layers is used for each charge network and energy network. L2 regularization and dropout \cite{srivastava2014dropout} on last layer were used to prevent overfitting with a dropout probability of 0.3. We chose a softplus function as the non-linear activation function after extensive experimentation, and used the Adaptive moment quasi-Newton solver (Adam)\cite{kingma2014adam} to fix the weights of the network. The test sets were separated from training data by choosing a random 20\% of molecules at the outset which were kept independent throughout. Besides water we will present calculations from molecules strictly absent from either the training or test set.

\begin{figure}
    \centering
    \includegraphics[width=0.45\textwidth]{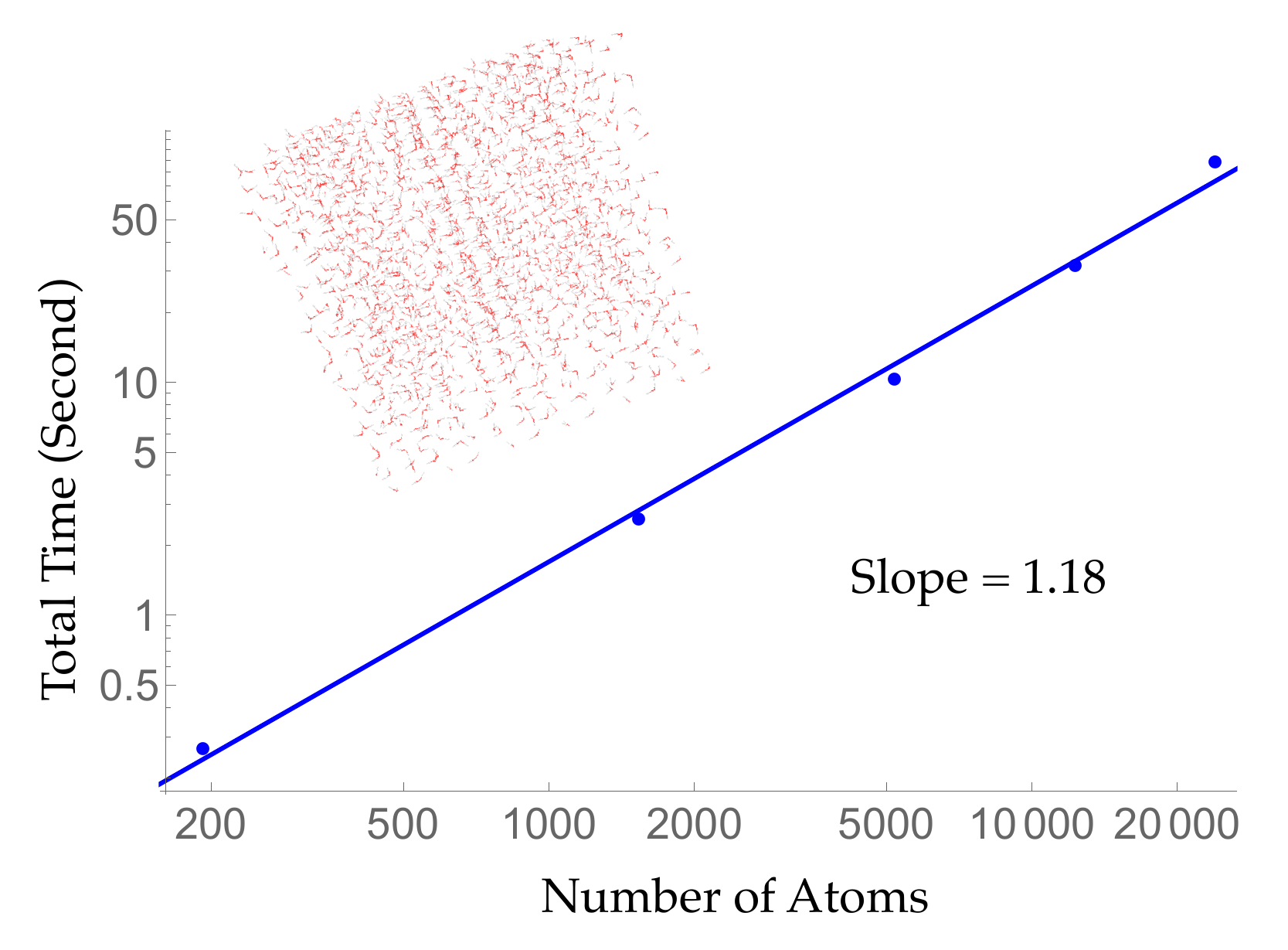}
    \caption{Aperiodic timings of an energy, charge, force call for cubic water clusters at a density of 1 gm/cm$^3$. The largest $\sim60$ Angstrom cube is 4x larger than the electrostatic cutoff. The slope of a log-log version of this curve is near unity, indicating the wall-time scaling of TensorMol.}
    \label{fig:timings}
\end{figure}

\indent     To obtain scalable efficiency TensorMol uses neighborlists within cutoffs. This allows double precision energy, charge, force calculations of up to 24,000 atoms to execute in less than 90 seconds on a 2015 Intel i7 2.5GHz MacBook pro (Fig. \ref{fig:timings}). Periodic evaluations are achieved by tessellation of a unit cell with summation of energies for atoms within the cell. This results in  roughly a factor of three overhead in computational time. Speedups are obtained automatically for computers with GPUs or single-precision calculations. 

\section{\label{sec:level1}Results}
\begin{figure}
    \centering
    \includegraphics[width=0.42\textwidth]{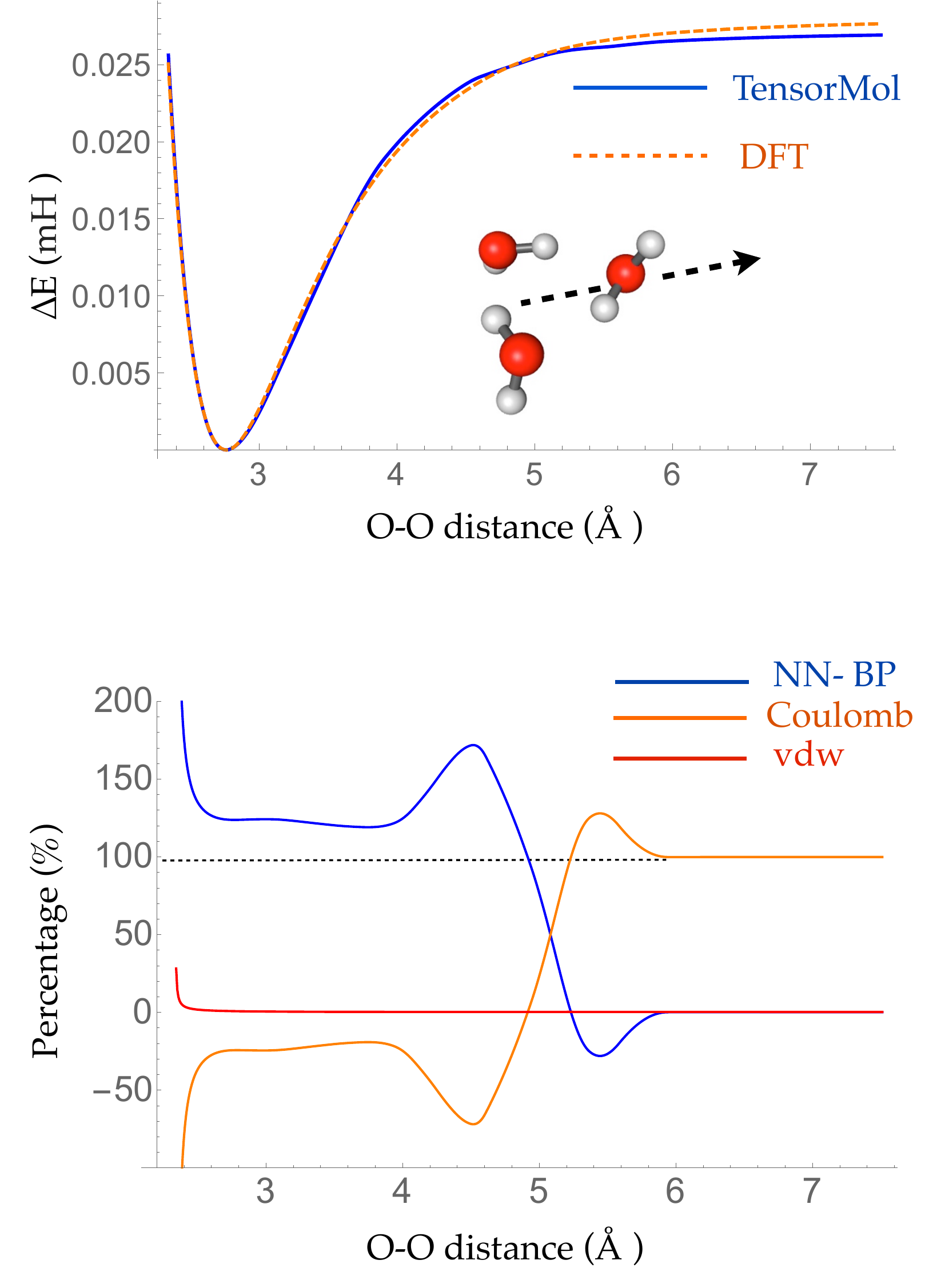}
    \caption{Left panel: PES of water trimer when one water is pulled away from the other two. Right panel: Percentage contribution of binding energy between the water that is pulled away and the other two water from Behler-Parrinello atom-wise energy, electrostatic energy and van-der waals energy. Behler-Parrinello atom-wise energy contribute to most of the binding energy at the short range and electrostatic energy is the dominant contribution at long range.}
    \label{fig:trimer}
\end{figure}

\begin{figure}
    \centering
    \includegraphics[width=0.42\textwidth]{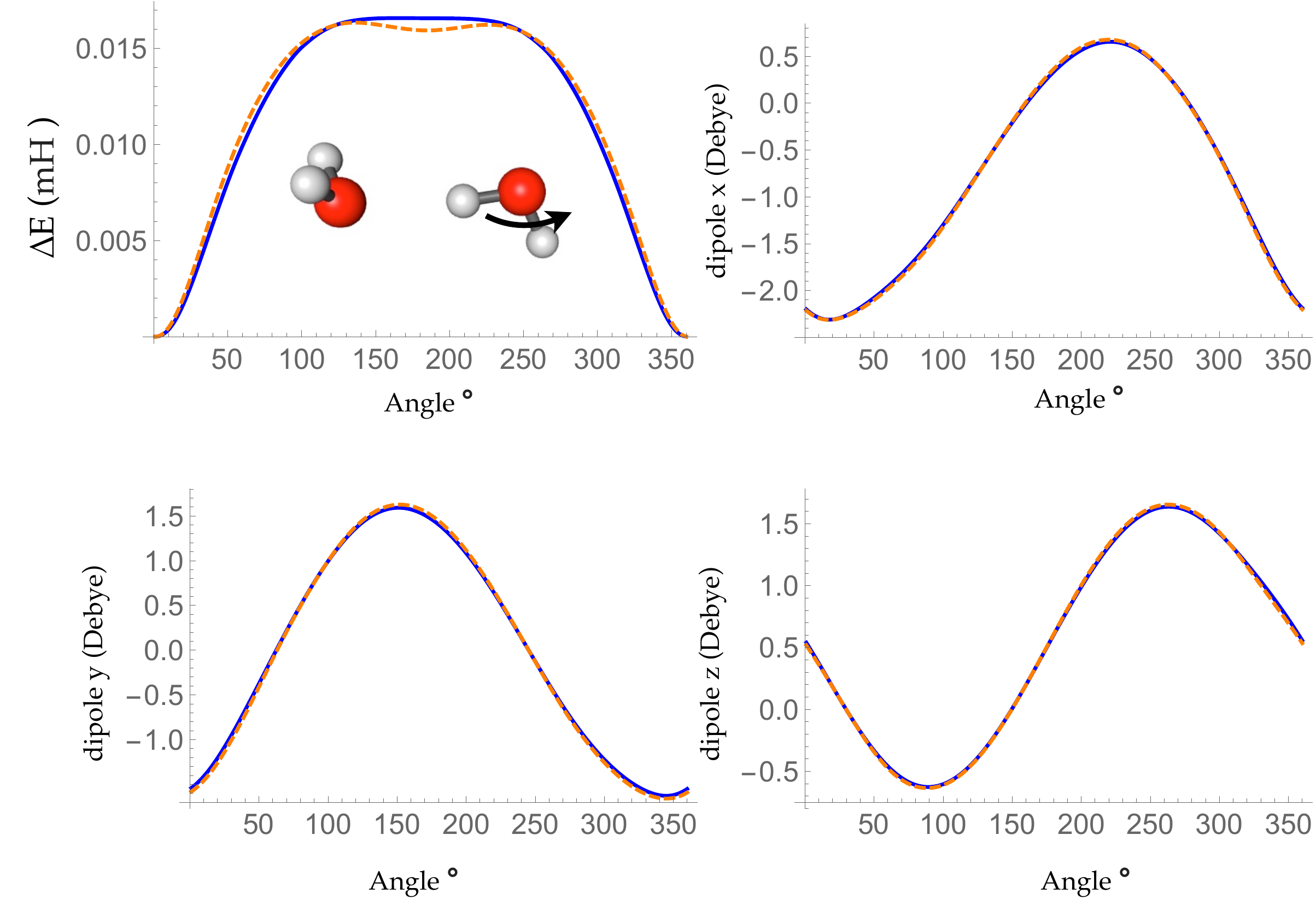}
    \caption{Top left panel: PES of breaking a hydrogen bond between two water by rotating one water around O-H bond. Top right, bottom left and bottom right panels: change of x, y, z, dipole component during the rotation, respectively}
    \label{fig:dimer}
\end{figure}

The root mean square error (RMSE) on the independent test set of the energy is 0.054 kcal/mol per atom and the RMSE of the force is 0.49 kcal/mol/\AA. The left panel of figure \ref{fig:trimer} plots the potential energy surface (PES) of a water trimer when one of the water is pulled away from the other two. One can see our neural network PES is not only in good agreement with the PES of target method but also smooth. To achieve this we use a variation of the soft-plus neuron rather than the rectified linear units which are popular in computer science. The latter train more efficiently, but produce discontinuous forces. \\
\indent     The right panel shows the fraction of each of the three energy components in equation 1 to the binding energy along the trimer dissociation coordinate.  At short range, most of the binding energy is contributed by the N-body neural network potential. When the distance between the monomer and the dimer approach the cutoff distance of the neural network, the contribution of neural network potential starts to decrease and the contribution of electrostatic potential increases. After 6\ \AA\ where the neural network symmetry functions on the atoms in the monomer have no contribution from the dimer, the neural network force drops smoothly to zero and the electrostatic interaction dominates. The small difference in the energy at 7\ \AA\ is due to the difference between the Madelung energy given by the learned charges, and the genuine physical cohesive force at this distance. The dimer and monomer are beyond the symmetry function sensory radius, and so the charges are constant in this region. Future iterations of the charge network will use local-field information to improve this region of the PES. The learned inductive charges are of high quality considering their linear scaling cost. Figure \ref{fig:dimer} shows the PES and dipole change of a water dimer when the hydrogen bond is broken by rotating the OH bond. Both the PES and dipole change fit well with the DFT results. \\

\begin{figure}
    \centering
    \includegraphics[width=0.42\textwidth]{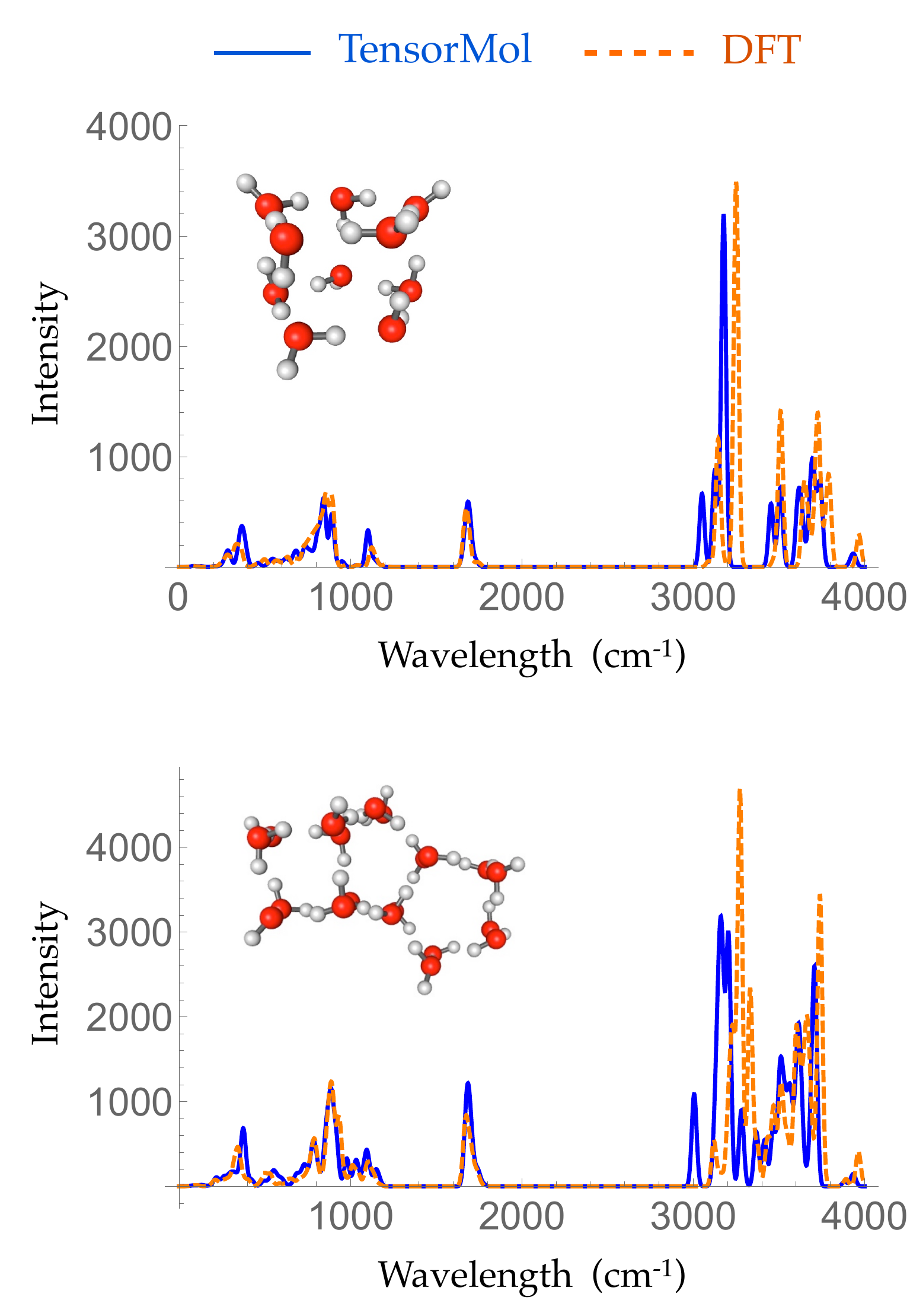}
    \caption{Simulated harmonic IR spectrum of 10 water cluster (left panel) and 20 water cluster (right panel) generated by WB97X-D/6-311G**(dashed orange line) and TensorMol force field (solid blue line).}
    \label{fig:waterir}
\end{figure}

Given the increased dimension of the Hessian, it is naturally a more stringent test to reproduce forces and infrared spectra than it is to simply produce energies. The left panel and right panel of figure \ref{fig:waterir} show the optimized geometries and IR spectra of a 10 water cluster and 20 water cluster generated with our force field and DFT, respectively. Each method uses its own equilibrium geometry, so this also tests TensorMol-0.1's reproduction of non-covalent geometry. The RMSE of the distance matrix between DFT optimized geometry and TensorMol optimized geometry are 0.062\ \AA\ for the 10 water cluster and 0.180\ \AA\ for the 20 water cluster. Our force field quantitatively reproduces the DFT IR both in terms of frequencies and intensities, especially for the water bend modes and inter-monomer modes. The Mean Absolute Error (MAE) of frequencies at that those two regions are 33.2 cm\textsuperscript{-1} for the 10 water cluster and 16.2 cm\textsuperscript{-1} for the 20 water cluster. The error is slightly larger at  water OH stretching region with a MAE of 34.2 cm\textsuperscript{-1} and 13.1 cm\textsuperscript{-1}, respectively. This accuracy is comparable to high quality polarizable water force fields\cite{medders2015representation}.\\

\begin{table}
\caption{\label{tab:table1} Training details and test RMSE of each learning target. The unit of energy RMSE, gradient RMSE and dipole RMSE is kcal/mol per atom, kcal/mol/\AA\ per atom and Debye per atom, respectively.}
\begin{ruledtabular}
\begin{tabular}{lcc}
&Water Network&Chemsipider Network\\
\hline
Num of training case & 370844 & 2979162\\
Training time (days)\footnote{Training was done on single Nvidia K40 GPU}& 3  & 10 \\
Energy RMSE& 0.054 & 0.24\\
Gradient RMSE& 0.49 & 2.4\\
Dipole RMSE& 0.0082  & 0.024\\
\end{tabular}
\end{ruledtabular}
\end{table}

\begin{figure}
    \centering
    \includegraphics[width=0.42\textwidth]{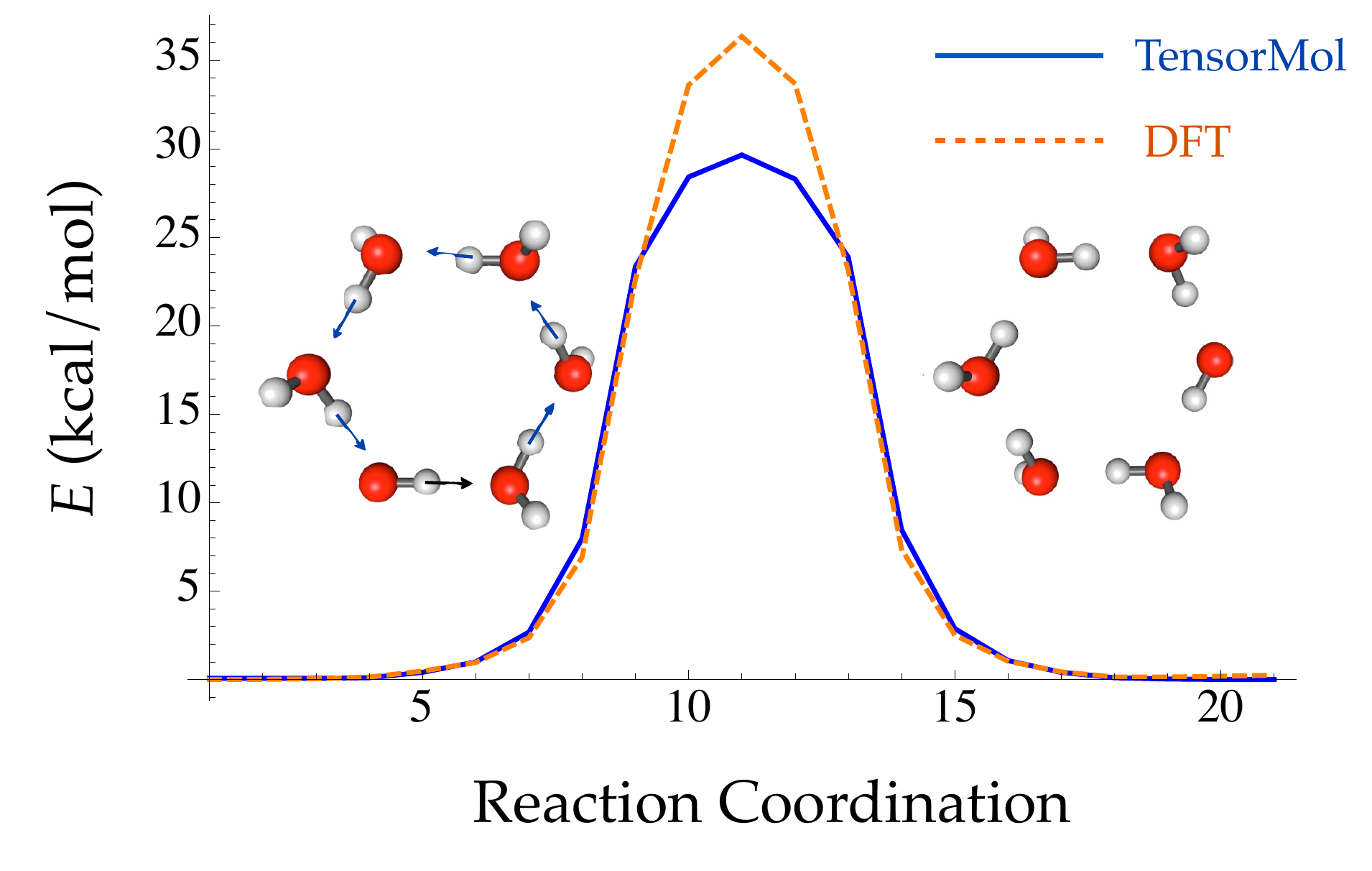}
    \caption{Reaction energy profile converged from a Nudged elastic band along the reaction coordinate of conservative proton transfer in a water hexamer cluster.}
    \label{fig:waterneb}
\end{figure}

Compared with traditional force fields, one major advantage of TensorMol is its reactivity. TensorMol is able to simulate a concerted proton transfer in a water hexamer, finding a minimum energy transition path. The PES's calculated by nudged elastic band (NEB) method \cite{henkelman2000climbing} with the TensorMol force field and DFT are shown in figure \ref{fig:waterneb}. The barrier height predicted by TensorMol is 36.3 kcal/mol, which is 6.7 kcal/mol lower than the prediction of DFT, which is remarkable considering the dearth of transition structures in the training data. Our sampling of molecular geometries uses a meta-dynamics procedure described elsewhere, so these proton transfers occur in the training data although extremely infrequently. 

\begin{figure}
    \centering
    \includegraphics[width=0.5\textwidth]{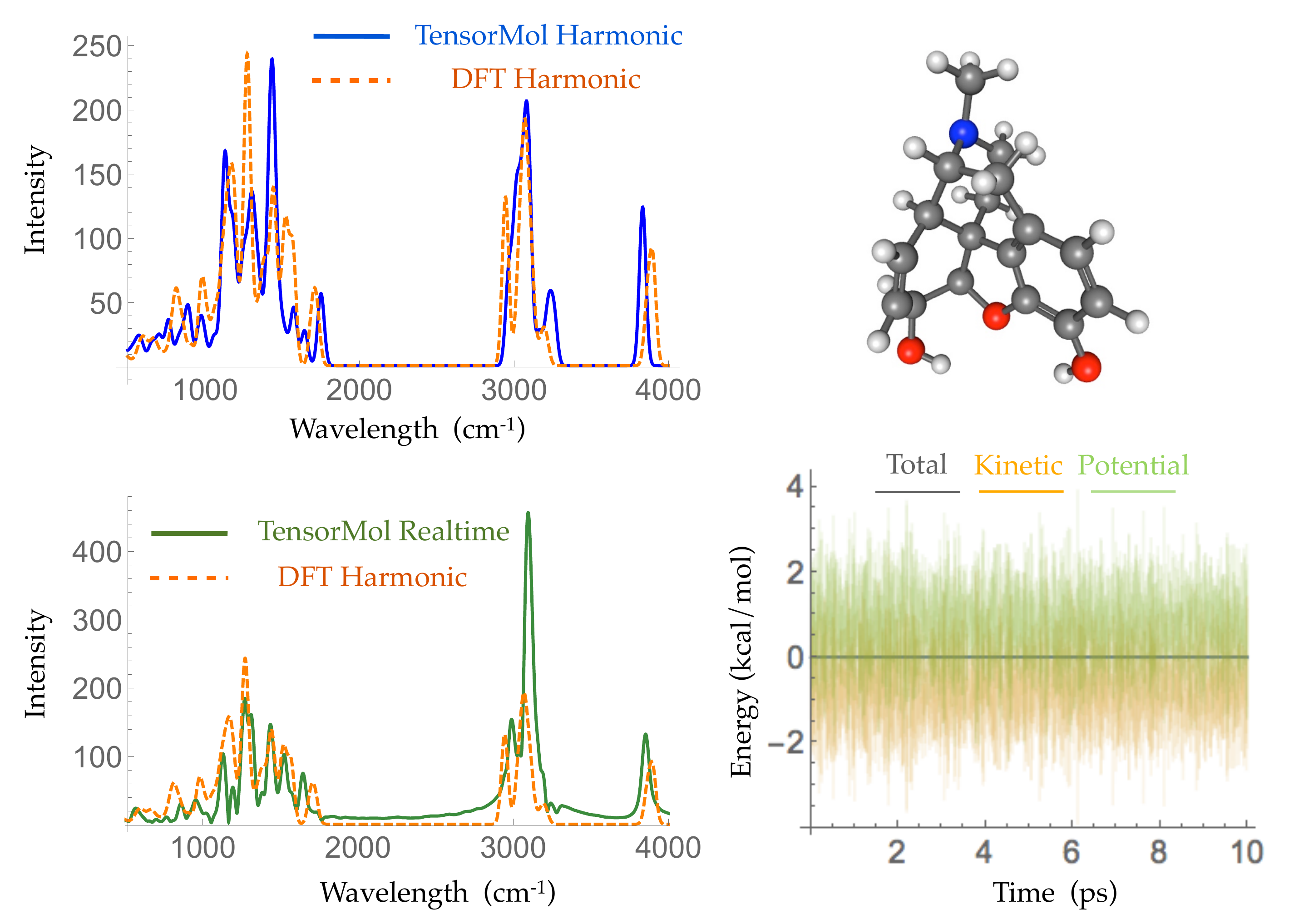}
    \caption{Morphine geometry that is optimized by TensorMol-0.1 (upper right panel) and its harmonic IR spectrum simulated by WB97X-D/6-311G**(dashed orange line) and TensorMol force field (solid blue line) (upper left panel). Lower panels show TensorMol's real-time IR spectrum vs. DFT (left) and the conservation of energy maintained by the smoothness of the energy (right).}
    \label{fig:morphine}
\end{figure}

\begin{figure}
    \centering
    \includegraphics[width=0.5\textwidth]{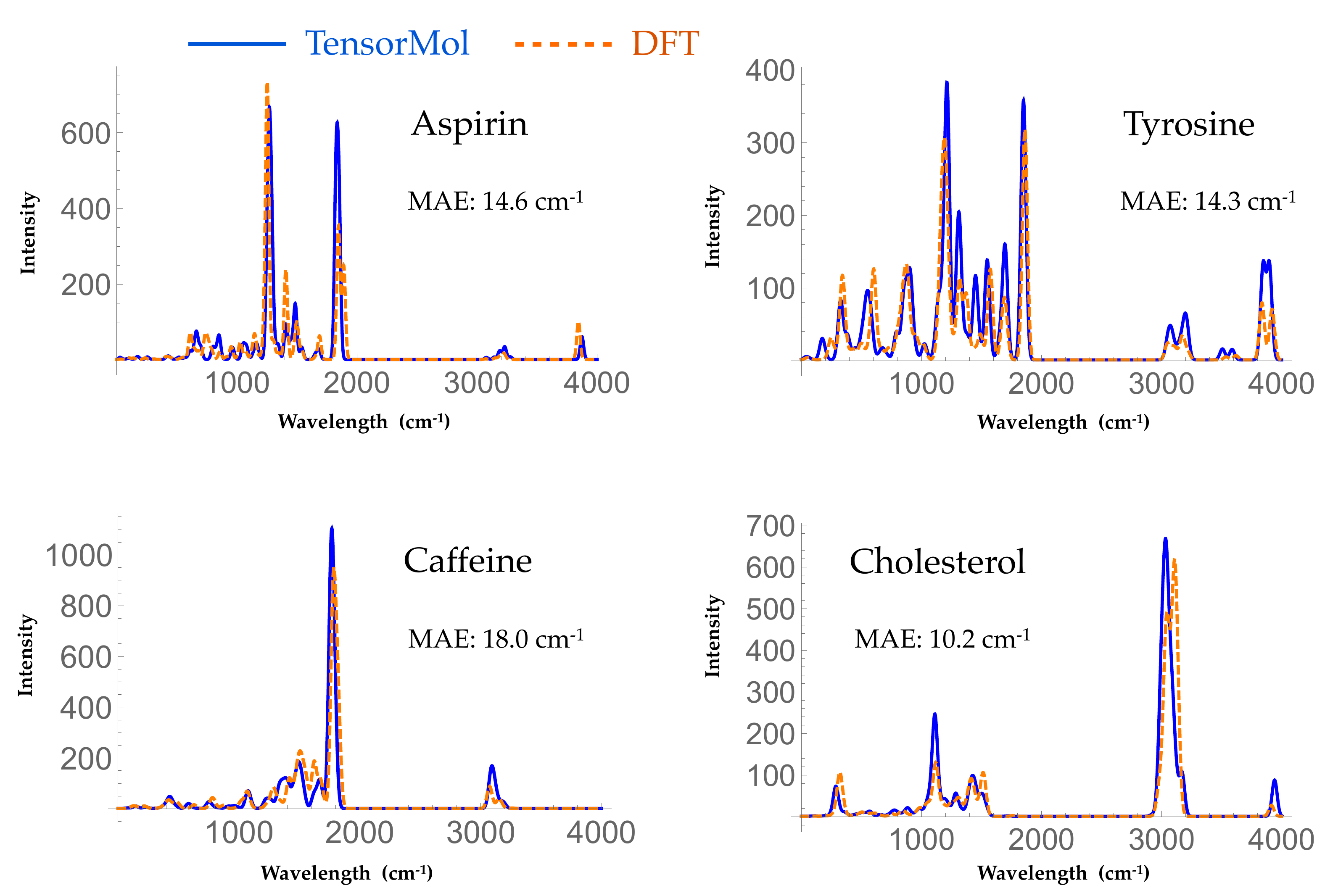}
    \caption{Harmonic IR spectrum of four different molecules simulated by WB97X-D/6-311G**(dashed orange line) and TensorMol-0.1. All the molecules are not included in the training set.}
    \label{fig:chemspider}
\end{figure}

\begin{figure}
    \centering
    \includegraphics[width=0.5\textwidth]{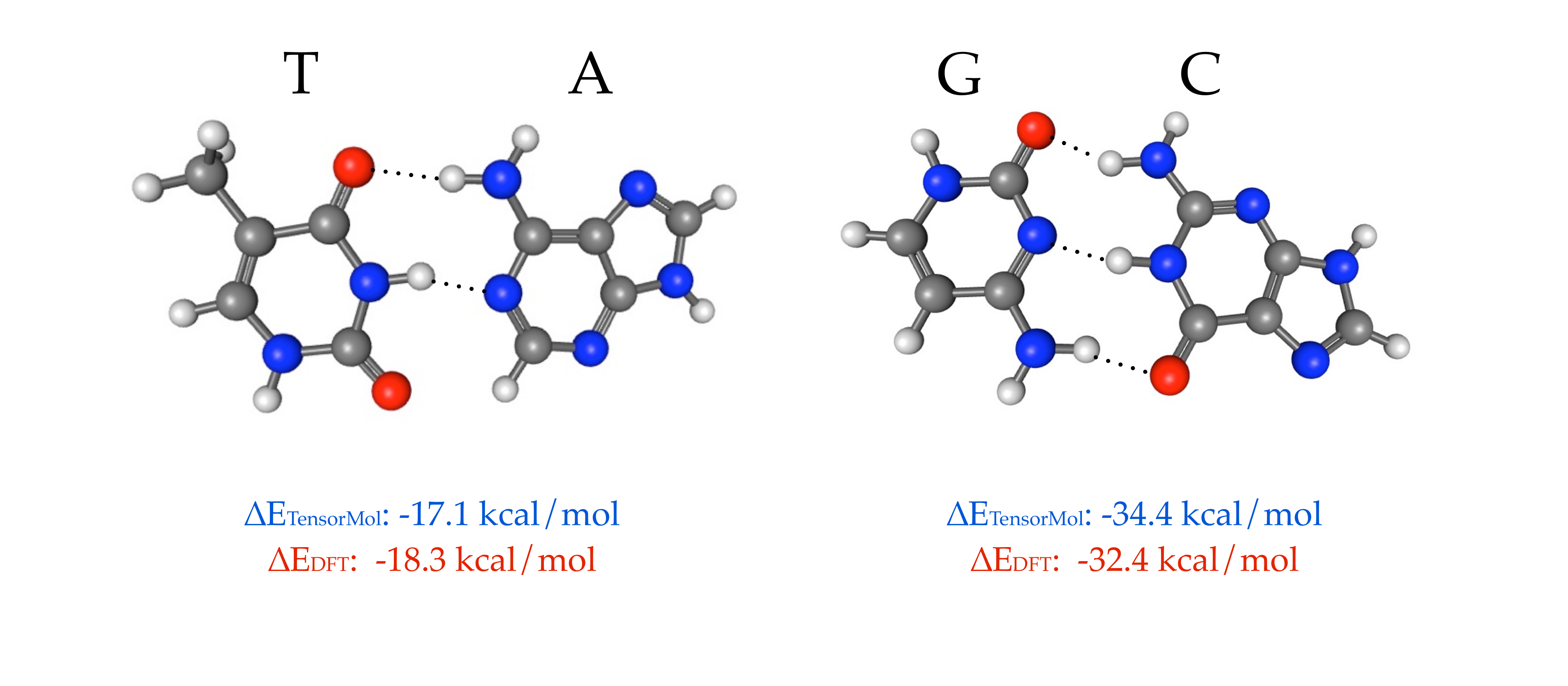}
    \caption{Binding energy between the DNA base pairs vs. $\omega$B97x-D with methods at their optimized geometries. The difference between DFT and TensorMol binding energy is < 2 kcal/mol.}
    \label{fig:DNA}
\end{figure}

\begin{figure}
    \centering
    \includegraphics[width=0.45\textwidth]{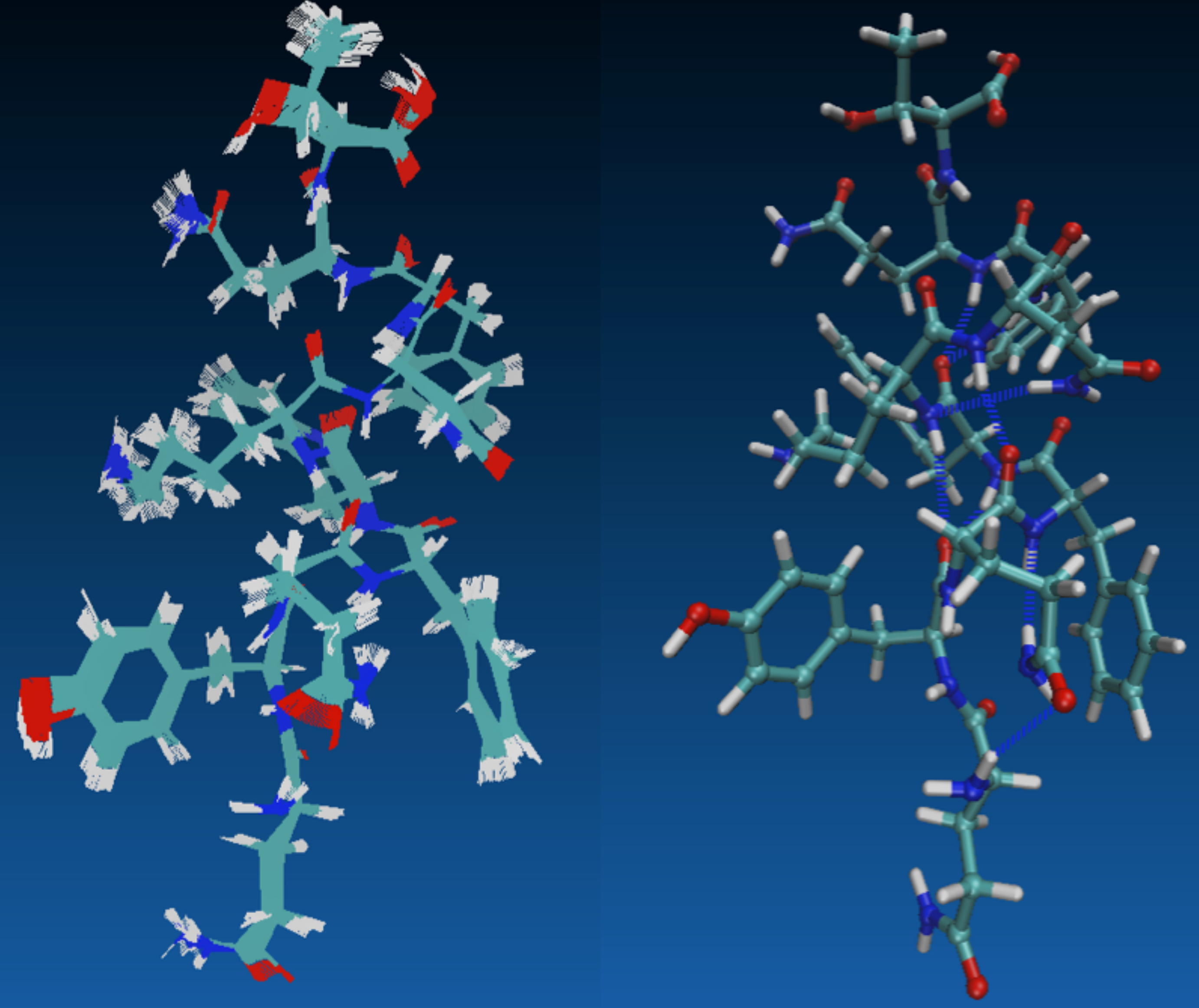}
    \caption{Left panel shows samples from a 1 picosecond NVT (Nos\'e) trajectory of solvated 2MZX at 300K simulated by our TensorMol force field in explicit water. Right panel is the NMR structure of 2MZX from the PDB database.}
    \label{fig:protein}
\end{figure}

Encouraged by our water results, we developed a force field with applicability across the chemical space spanned by CNOH. The Chemspider dataset that we used to train our force field covers a vast chemical space containing 15 thousand different molecules and 3 millions geometries. The geometries are generated using a meta-dynamics procedure\cite{barducci2008well}, which ensures that each new geometry is a fresh part of chemical space, energies up to 400$k_bT$ are sampled in the data. We describe the details of this meta-dynamics sampling algorithm, which we have found vital to achieving robust and transferrable force-fields elsewhere. The diversity of structures makes learning the chemspider dataset a much harder task for neural networks, the test set RMSE of energy is 0.24 kcal/mol per atom and RMSE of force is 2.4 kcal/mol per atom. More importantly, the model usefully reproduces several elements of molecular structure at and away from equilibrium for molecules outside its training set. It robustly optimizes the geometries of typical organic molecules to structures that match DFT well, and yields infrared frequencies and intensities in good agreement with ab-initio calculations. It is a black-box method which does not rely on any atom type, connectivity, etc as one would need to specify in a traditional classical force-field. The few proteins we have examined remain stable and near their experimental structures when optimized or propagated at room temperature using the TensorMol-0.1 force field.\\
\indent Morphine is not included in our training set. The left panel of figure \ref{fig:morphine} shows the geometry of morphine that is optimized with our force field. The RMSE of bond lengths predicted by our forcefield is 0.0067\ \AA\ and the RMSE of angles is 1.04 degrees compared with the source DFT model chemistry. The right panel plots the harmonic IR spectra generated by each method at their respective optimized geometries. One can see IR spectrum from our force field is in good agreement with the DFT IR spectrum. The MAE of our force field frequencies is 13.7 cm\textsuperscript{-1} compared with DFT frequencies. Figure \ref{fig:chemspider} shows comparisons of IR spectrum that are generated from these two methods for aspirin, typrosine, caffeine and cholesterol. All these four molecules are not included in the training set. The MAE of the frequencies predicted by our field is less than 20 cm\textsuperscript{-1} for all the four molecules compared with target DFT frequencies. The concept of a chemical bond and force constant are not enforced in any way, yet good agreement with DFT is obtained at a tiny fraction of the original cost. \\
\indent    Traditional harmonic vibrational spectra require quadratic computational effort, which works against the speed advantage of a NNMC. For large systems one can use the molecular dynamics functionality of TensorMol to simulate infrared spectra, Fourier transforming the dipole-dipole correlation function of conservative Newtonian dynamics whose cost grows linearly with the size of the system. The lower left panel of Figure \ref{fig:morphine} shows the same infrared spectrum produced by propagation in TensorMol-0.1, also showcasing the good energy conservation of TensorMol. Unlike a traditional force-field it's non-trivial to obtain smoothly differentiable NNMC's. 64-bit precision needs to be used the network cannot be made too flexible and smooth versions of typical rectified linear units need to be used. Our package can be used in this way to simulate IR of large systems with linear cost.\\
\indent     TensorMol-0.1 has a relatively simple treatment of electrostatic and Van-Der-Waals forces which we would like to augment in the future with a many-body dispersion scheme\cite{tkatchenko2012accurate}. However a main adantage of TensorMol-0.1's approach is its very low cost. No self-consistent polarization equation is solved even though the charges are inductive, and so it is easy to inexpensively calculate the electrostatic energies of even very large molecules. At shorter ranges, non-covalent interactions like hydrogen bonds are dealt with by the Behler-Parinello portion of the network. The Chemspider training data includes some examples of dimers and intra-molecular hydrogen bonds. To our surprise the treatment of inter-molecular interactions which were not targets for TensorMol-0.1 are satisfactory. Figure \ref{fig:DNA} shows the optimized geometries and binding energies of two DNA base pairs calculated by our force field. The target DFT method predicts a binding energy of 18.3 kcal/mol for the thymine-adenine (TA) pair and a binding energy of 32.4 kcal/mol for the guanine-cytosine (GC) pair. The prediction of our force field is 1.2 kcal/mol less for TA pair and 2.0 kcal/mol larger for GC pairs relative to DFT. \\ 
\indent     One holy grail of the field of Neural Network model chemistries is to simulate biological chemistry without QM-MM or bespoke force-fields. Protein simulation also demonstrates several important features of a neural network model chemistry: reasonable inter-molecular forces, stability, scalability and generalization far from small-molecule training data. TensorMol-0.1 was not trained on any peptide polymers and includes no biological data of any sort. To our pleasant surprise, even this first iteration of Neural Network model chemistry is accurate enough to perform rudimentary studies of small proteins. Figure \ref{fig:protein} shows geometries sampled from a 1 picosecond, periodic, 300K TensorMol dynamics NVT trajectory in explicit solvent. The initial structure (included in the supplement) was generated from the PDB structure 2MZX using OpenMM's automatic solvation and hydrogenation scripts\cite{OpenMM}, but includes nothing but atom coordinates. This short alpha-helix is stable, both in optimizations and dynamics, and the structures sampled during the dynamics superficially resemble the solution NMR structure. Traditional force fields will always be less expensive (by some prefactor) than NNMCs, yet the reactivity advantages of NNMCs and the ease of set up will probably lead to a rapid adoption of these methods in the biological community. 
\subsection{Discussion and Conclusions}
\indent     We have presented a transferable neural network model chemistry with long-range Coulombic physics, and a short-range n-Body potential, TensorMol-0.1. The model is integrated in a concise open-source python package which provides many of the types of simulation commonly used in chemistry. The method can be used to scan conformational and chemical space along the singlet neutral potential energy surface with high throughput and accuracy using nothing but atomic coordinates. TensorMol-0.1 is not the final iteration of Neural-Network model chemistry, although it shows just how easily DFT-quality predictions can be made by models with drastically lower cost. Inexpensive post-DFT corrections such as Many-Body Dispersion \cite{tkatchenko2012accurate} will become even more powerful when integrated with these potentials, opening the door to quantitative treatments of large systems. These methods will compete aggressively with DFT packages, and provide an interesting complement to QM-MM type simulations in the near future. \\
\indent     Methods like TensorMol-0.1 suggest the following directions for the future improvement of Neural-Network model chemistries:
\begin{itemize}
\item Generalize descriptors to encode other physical atom properties besides charge (spin, polarizability)
\item Develop accurate descriptors whose cost grows linearly with the number of elements treated
\item Extend the range of the N-Body embedding
\item Explore the hierarchy of physical detail between Force-Fields and semi-empirical electronic structure. 
\end{itemize}
These goals must be pursed alongside honest test suites with open programs and data. NNMCs that cannot produce forces and MD trajectories with a demonstrable and compelling cost advantage over DFT should only be pursued if they offer interesting qualitative insights. 

\begin{acknowledgments}
The authors gratefully acknowledge Notre Dame's College of Science for startup funding, Oak Ridge national laboratory for a grant of supercomputer resources and NVidia corporation.
\end{acknowledgments}

%


\begin{thebibliography}{96}%
\makeatletter
\providecommand \@ifxundefined [1]{%
 \@ifx{#1\undefined}
}%
\providecommand \@ifnum [1]{%
 \ifnum #1\expandafter \@firstoftwo
 \else \expandafter \@secondoftwo
 \fi
}%
\providecommand \@ifx [1]{%
 \ifx #1\expandafter \@firstoftwo
 \else \expandafter \@secondoftwo
 \fi
}%
\providecommand \natexlab [1]{#1}%
\providecommand \enquote  [1]{``#1''}%
\providecommand \bibnamefont  [1]{#1}%
\providecommand \bibfnamefont [1]{#1}%
\providecommand \citenamefont [1]{#1}%
\providecommand \href@noop [0]{\@secondoftwo}%
\providecommand \href [0]{\begingroup \@sanitize@url \@href}%
\providecommand \@href[1]{\@@startlink{#1}\@@href}%
\providecommand \@@href[1]{\endgroup#1\@@endlink}%
\providecommand \@sanitize@url [0]{\catcode `\\12\catcode `\$12\catcode
  `\&12\catcode `\#12\catcode `\^12\catcode `\_12\catcode `\%12\relax}%
\providecommand \@@startlink[1]{}%
\providecommand \@@endlink[0]{}%
\providecommand \url  [0]{\begingroup\@sanitize@url \@url }%
\providecommand \@url [1]{\endgroup\@href {#1}{\urlprefix }}%
\providecommand \urlprefix  [0]{URL }%
\providecommand \Eprint [0]{\href }%
\providecommand \doibase [0]{http://dx.doi.org/}%
\providecommand \selectlanguage [0]{\@gobble}%
\providecommand \bibinfo  [0]{\@secondoftwo}%
\providecommand \bibfield  [0]{\@secondoftwo}%
\providecommand \translation [1]{[#1]}%
\providecommand \BibitemOpen [0]{}%
\providecommand \bibitemStop [0]{}%
\providecommand \bibitemNoStop [0]{.\EOS\space}%
\providecommand \EOS [0]{\spacefactor3000\relax}%
\providecommand \BibitemShut  [1]{\csname bibitem#1\endcsname}%
\let\auto@bib@innerbib\@empty
\bibitem [{\citenamefont {Snyder}\ \emph {et~al.}(2013)\citenamefont {Snyder},
  \citenamefont {Rupp}, \citenamefont {Hansen}, \citenamefont {Blooston},
  \citenamefont {M\"u{}ller},\ and\ \citenamefont {Burke}}]{Snyder:2013aa}%
  \BibitemOpen
  \bibfield  {author} {\bibinfo {author} {\bibfnamefont {J.~C.}\ \bibnamefont
  {Snyder}}, \bibinfo {author} {\bibfnamefont {M.}~\bibnamefont {Rupp}},
  \bibinfo {author} {\bibfnamefont {K.}~\bibnamefont {Hansen}}, \bibinfo
  {author} {\bibfnamefont {L.}~\bibnamefont {Blooston}}, \bibinfo {author}
  {\bibfnamefont {K.-R.}\ \bibnamefont {M\"u{}ller}}, \ and\ \bibinfo {author}
  {\bibfnamefont {K.}~\bibnamefont {Burke}},\ }\href {\doibase
  http://dx.doi.org/10.1063/1.4834075} {\bibfield  {journal} {\bibinfo
  {journal} {J. Chem. Phys.}\ }\textbf {\bibinfo {volume} {139}},\ \bibinfo
  {pages} {224104} (\bibinfo {year} {2013})}\BibitemShut {NoStop}%
\bibitem [{\citenamefont {Brockherde}\ \emph {et~al.}(2017)\citenamefont
  {Brockherde}, \citenamefont {Vogt}, \citenamefont {Li}, \citenamefont
  {Tuckerman}, \citenamefont {Burke},\ and\ \citenamefont
  {M{\"u}ller}}]{brockherde2017bypassing}%
  \BibitemOpen
  \bibfield  {author} {\bibinfo {author} {\bibfnamefont {F.}~\bibnamefont
  {Brockherde}}, \bibinfo {author} {\bibfnamefont {L.}~\bibnamefont {Vogt}},
  \bibinfo {author} {\bibfnamefont {L.}~\bibnamefont {Li}}, \bibinfo {author}
  {\bibfnamefont {M.~E.}\ \bibnamefont {Tuckerman}}, \bibinfo {author}
  {\bibfnamefont {K.}~\bibnamefont {Burke}}, \ and\ \bibinfo {author}
  {\bibfnamefont {K.-R.}\ \bibnamefont {M{\"u}ller}},\ }\href@noop {}
  {\bibfield  {journal} {\bibinfo  {journal} {Nat. Commun.}\ }\textbf {\bibinfo
  {volume} {8}},\ \bibinfo {pages} {872} (\bibinfo {year} {2017})}\BibitemShut
  {NoStop}%
\bibitem [{\citenamefont {Snyder}\ \emph {et~al.}(2012)\citenamefont {Snyder},
  \citenamefont {Rupp}, \citenamefont {Hansen}, \citenamefont {M\"u{}ller},\
  and\ \citenamefont {Burke}}]{snyder2012finding}%
  \BibitemOpen
  \bibfield  {author} {\bibinfo {author} {\bibfnamefont {J.~C.}\ \bibnamefont
  {Snyder}}, \bibinfo {author} {\bibfnamefont {M.}~\bibnamefont {Rupp}},
  \bibinfo {author} {\bibfnamefont {K.}~\bibnamefont {Hansen}}, \bibinfo
  {author} {\bibfnamefont {K.-R.}\ \bibnamefont {M\"u{}ller}}, \ and\ \bibinfo
  {author} {\bibfnamefont {K.}~\bibnamefont {Burke}},\ }\href@noop {}
  {\bibfield  {journal} {\bibinfo  {journal} {Phys. Rev. Lett.}\ }\textbf
  {\bibinfo {volume} {108}},\ \bibinfo {pages} {253002} (\bibinfo {year}
  {2012})}\BibitemShut {NoStop}%
\bibitem [{\citenamefont {Li}\ \emph {et~al.}(2016{\natexlab{a}})\citenamefont
  {Li}, \citenamefont {Snyder}, \citenamefont {Pelaschier}, \citenamefont
  {Huang}, \citenamefont {Niranjan}, \citenamefont {Duncan}, \citenamefont
  {Rupp}, \citenamefont {M{\"u}ller},\ and\ \citenamefont
  {Burke}}]{li2016understanding}%
  \BibitemOpen
  \bibfield  {author} {\bibinfo {author} {\bibfnamefont {L.}~\bibnamefont
  {Li}}, \bibinfo {author} {\bibfnamefont {J.~C.}\ \bibnamefont {Snyder}},
  \bibinfo {author} {\bibfnamefont {I.~M.}\ \bibnamefont {Pelaschier}},
  \bibinfo {author} {\bibfnamefont {J.}~\bibnamefont {Huang}}, \bibinfo
  {author} {\bibfnamefont {U.-N.}\ \bibnamefont {Niranjan}}, \bibinfo {author}
  {\bibfnamefont {P.}~\bibnamefont {Duncan}}, \bibinfo {author} {\bibfnamefont
  {M.}~\bibnamefont {Rupp}}, \bibinfo {author} {\bibfnamefont {K.-R.}\
  \bibnamefont {M{\"u}ller}}, \ and\ \bibinfo {author} {\bibfnamefont
  {K.}~\bibnamefont {Burke}},\ }\href@noop {} {\bibfield  {journal} {\bibinfo
  {journal} {Int. J. Quantum Chem.}\ }\textbf {\bibinfo {volume} {116}},\
  \bibinfo {pages} {819} (\bibinfo {year} {2016}{\natexlab{a}})}\BibitemShut
  {NoStop}%
\bibitem [{\citenamefont {Li}\ \emph {et~al.}(2016{\natexlab{b}})\citenamefont
  {Li}, \citenamefont {Baker}, \citenamefont {White}, \citenamefont {Burke}
  \emph {et~al.}}]{li2016pure}%
  \BibitemOpen
  \bibfield  {author} {\bibinfo {author} {\bibfnamefont {L.}~\bibnamefont
  {Li}}, \bibinfo {author} {\bibfnamefont {T.~E.}\ \bibnamefont {Baker}},
  \bibinfo {author} {\bibfnamefont {S.~R.}\ \bibnamefont {White}}, \bibinfo
  {author} {\bibfnamefont {K.}~\bibnamefont {Burke}},  \emph {et~al.},\
  }\href@noop {} {\bibfield  {journal} {\bibinfo  {journal} {Phys. Rev. B}\
  }\textbf {\bibinfo {volume} {94}},\ \bibinfo {pages} {245129} (\bibinfo
  {year} {2016}{\natexlab{b}})}\BibitemShut {NoStop}%
\bibitem [{\citenamefont {Vu}\ \emph {et~al.}(2015)\citenamefont {Vu},
  \citenamefont {Snyder}, \citenamefont {Li}, \citenamefont {Rupp},
  \citenamefont {Chen}, \citenamefont {Khelif}, \citenamefont {M{\"u}ller},\
  and\ \citenamefont {Burke}}]{vu2015understanding}%
  \BibitemOpen
  \bibfield  {author} {\bibinfo {author} {\bibfnamefont {K.}~\bibnamefont
  {Vu}}, \bibinfo {author} {\bibfnamefont {J.~C.}\ \bibnamefont {Snyder}},
  \bibinfo {author} {\bibfnamefont {L.}~\bibnamefont {Li}}, \bibinfo {author}
  {\bibfnamefont {M.}~\bibnamefont {Rupp}}, \bibinfo {author} {\bibfnamefont
  {B.~F.}\ \bibnamefont {Chen}}, \bibinfo {author} {\bibfnamefont
  {T.}~\bibnamefont {Khelif}}, \bibinfo {author} {\bibfnamefont {K.-R.}\
  \bibnamefont {M{\"u}ller}}, \ and\ \bibinfo {author} {\bibfnamefont
  {K.}~\bibnamefont {Burke}},\ }\href@noop {} {\bibfield  {journal} {\bibinfo
  {journal} {Int. J. Quantum Chem.}\ }\textbf {\bibinfo {volume} {115}},\
  \bibinfo {pages} {1115} (\bibinfo {year} {2015})}\BibitemShut {NoStop}%
\bibitem [{\citenamefont {Handley}\ and\ \citenamefont
  {Popelier}(2010)}]{handley2010potential}%
  \BibitemOpen
  \bibfield  {author} {\bibinfo {author} {\bibfnamefont {C.~M.}\ \bibnamefont
  {Handley}}\ and\ \bibinfo {author} {\bibfnamefont {P.~L.}\ \bibnamefont
  {Popelier}},\ }\href@noop {} {\bibfield  {journal} {\bibinfo  {journal} {J.
  Phys. Chem. A}\ }\textbf {\bibinfo {volume} {114}},\ \bibinfo {pages} {3371}
  (\bibinfo {year} {2010})}\BibitemShut {NoStop}%
\bibitem [{\citenamefont {Chmiela}\ \emph {et~al.}(2017)\citenamefont
  {Chmiela}, \citenamefont {Tkatchenko}, \citenamefont {Sauceda}, \citenamefont
  {Poltavsky}, \citenamefont {Sch{\"u}tt},\ and\ \citenamefont
  {M{\"u}ller}}]{chmiela2017machine}%
  \BibitemOpen
  \bibfield  {author} {\bibinfo {author} {\bibfnamefont {S.}~\bibnamefont
  {Chmiela}}, \bibinfo {author} {\bibfnamefont {A.}~\bibnamefont {Tkatchenko}},
  \bibinfo {author} {\bibfnamefont {H.~E.}\ \bibnamefont {Sauceda}}, \bibinfo
  {author} {\bibfnamefont {I.}~\bibnamefont {Poltavsky}}, \bibinfo {author}
  {\bibfnamefont {K.~T.}\ \bibnamefont {Sch{\"u}tt}}, \ and\ \bibinfo {author}
  {\bibfnamefont {K.-R.}\ \bibnamefont {M{\"u}ller}},\ }\href@noop {}
  {\bibfield  {journal} {\bibinfo  {journal} {Sci. Adv.}\ }\textbf {\bibinfo
  {volume} {3}},\ \bibinfo {pages} {e1603015} (\bibinfo {year}
  {2017})}\BibitemShut {NoStop}%
\bibitem [{\citenamefont {Behler}(2011)}]{behler2011neural}%
  \BibitemOpen
  \bibfield  {author} {\bibinfo {author} {\bibfnamefont {J.}~\bibnamefont
  {Behler}},\ }\href@noop {} {\bibfield  {journal} {\bibinfo  {journal} {Phys.
  Chem. Chem. Phys.}\ }\textbf {\bibinfo {volume} {13}},\ \bibinfo {pages}
  {17930} (\bibinfo {year} {2011})}\BibitemShut {NoStop}%
\bibitem [{\citenamefont {Behler}\ and\ \citenamefont
  {Parrinello}(2007)}]{behler2007generalized}%
  \BibitemOpen
  \bibfield  {author} {\bibinfo {author} {\bibfnamefont {J.}~\bibnamefont
  {Behler}}\ and\ \bibinfo {author} {\bibfnamefont {M.}~\bibnamefont
  {Parrinello}},\ }\href@noop {} {\bibfield  {journal} {\bibinfo  {journal}
  {Phys. Rev. Lett.}\ }\textbf {\bibinfo {volume} {98}},\ \bibinfo {pages}
  {146401} (\bibinfo {year} {2007})}\BibitemShut {NoStop}%
\bibitem [{\citenamefont {Shakouri}\ \emph {et~al.}(2017)\citenamefont
  {Shakouri}, \citenamefont {Behler}, \citenamefont {Meyer},\ and\
  \citenamefont {Kroes}}]{shakouri2017accurate}%
  \BibitemOpen
  \bibfield  {author} {\bibinfo {author} {\bibfnamefont {K.}~\bibnamefont
  {Shakouri}}, \bibinfo {author} {\bibfnamefont {J.}~\bibnamefont {Behler}},
  \bibinfo {author} {\bibfnamefont {J.}~\bibnamefont {Meyer}}, \ and\ \bibinfo
  {author} {\bibfnamefont {G.-J.}\ \bibnamefont {Kroes}},\ }\href@noop {}
  {\bibfield  {journal} {\bibinfo  {journal} {J. Phys. Chem. Lett.}\ }\textbf
  {\bibinfo {volume} {8}},\ \bibinfo {pages} {2131} (\bibinfo {year}
  {2017})}\BibitemShut {NoStop}%
\bibitem [{\citenamefont {Behler}(2017)}]{behler2017first}%
  \BibitemOpen
  \bibfield  {author} {\bibinfo {author} {\bibfnamefont {J.}~\bibnamefont
  {Behler}},\ }\href@noop {} {\bibfield  {journal} {\bibinfo  {journal} {Angew.
  Chem., Int. Ed.}\ }\textbf {\bibinfo {volume} {56}},\ \bibinfo {pages}
  {12828} (\bibinfo {year} {2017})}\BibitemShut {NoStop}%
\bibitem [{\citenamefont {Han}\ \emph {et~al.}(2017)\citenamefont {Han},
  \citenamefont {Zhang}, \citenamefont {Car} \emph {et~al.}}]{han2017deep}%
  \BibitemOpen
  \bibfield  {author} {\bibinfo {author} {\bibfnamefont {J.}~\bibnamefont
  {Han}}, \bibinfo {author} {\bibfnamefont {L.}~\bibnamefont {Zhang}}, \bibinfo
  {author} {\bibfnamefont {R.}~\bibnamefont {Car}},  \emph {et~al.},\
  }\href@noop {} {\bibfield  {journal} {\bibinfo  {journal} {arXiv preprint
  arXiv:1707.01478}\ } (\bibinfo {year} {2017})}\BibitemShut {NoStop}%
\bibitem [{\citenamefont {Yao}, \citenamefont {Herr},\ and\ \citenamefont
  {Parkhill}(2017)}]{yao2017many}%
  \BibitemOpen
  \bibfield  {author} {\bibinfo {author} {\bibfnamefont {K.}~\bibnamefont
  {Yao}}, \bibinfo {author} {\bibfnamefont {J.~E.}\ \bibnamefont {Herr}}, \
  and\ \bibinfo {author} {\bibfnamefont {J.}~\bibnamefont {Parkhill}},\
  }\href@noop {} {\bibfield  {journal} {\bibinfo  {journal} {J. Chem. Phys.}\
  }\textbf {\bibinfo {volume} {146}},\ \bibinfo {pages} {014106} (\bibinfo
  {year} {2017})}\BibitemShut {NoStop}%
\bibitem [{\citenamefont {Yao}\ and\ \citenamefont
  {Parkhill}(2016)}]{yao2016kinetic}%
  \BibitemOpen
  \bibfield  {author} {\bibinfo {author} {\bibfnamefont {K.}~\bibnamefont
  {Yao}}\ and\ \bibinfo {author} {\bibfnamefont {J.}~\bibnamefont {Parkhill}},\
  }\href@noop {} {\bibfield  {journal} {\bibinfo  {journal} {J. Chem. Theory
  Comput.}\ }\textbf {\bibinfo {volume} {12}},\ \bibinfo {pages} {1139}
  (\bibinfo {year} {2016})}\BibitemShut {NoStop}%
\bibitem [{\citenamefont {Yao}\ \emph {et~al.}(2017)\citenamefont {Yao},
  \citenamefont {Herr}, \citenamefont {Brown},\ and\ \citenamefont
  {Parkhill}}]{yao2017intrinsic}%
  \BibitemOpen
  \bibfield  {author} {\bibinfo {author} {\bibfnamefont {K.}~\bibnamefont
  {Yao}}, \bibinfo {author} {\bibfnamefont {J.~E.}\ \bibnamefont {Herr}},
  \bibinfo {author} {\bibfnamefont {S.~N.}\ \bibnamefont {Brown}}, \ and\
  \bibinfo {author} {\bibfnamefont {J.}~\bibnamefont {Parkhill}},\ }\href@noop
  {} {\bibfield  {journal} {\bibinfo  {journal} {J. Phys. Chem. Lett.}\ }
  (\bibinfo {year} {2017})}\BibitemShut {NoStop}%
\bibitem [{\citenamefont {Khaliullin}\ \emph {et~al.}(2011)\citenamefont
  {Khaliullin}, \citenamefont {Eshet}, \citenamefont {K{\"u}hne}, \citenamefont
  {Behler},\ and\ \citenamefont {Parrinello}}]{khaliullin2011nucleation}%
  \BibitemOpen
  \bibfield  {author} {\bibinfo {author} {\bibfnamefont {R.~Z.}\ \bibnamefont
  {Khaliullin}}, \bibinfo {author} {\bibfnamefont {H.}~\bibnamefont {Eshet}},
  \bibinfo {author} {\bibfnamefont {T.~D.}\ \bibnamefont {K{\"u}hne}}, \bibinfo
  {author} {\bibfnamefont {J.}~\bibnamefont {Behler}}, \ and\ \bibinfo {author}
  {\bibfnamefont {M.}~\bibnamefont {Parrinello}},\ }\href@noop {} {\bibfield
  {journal} {\bibinfo  {journal} {Nat. Mater.}\ }\textbf {\bibinfo {volume}
  {10}},\ \bibinfo {pages} {693} (\bibinfo {year} {2011})}\BibitemShut
  {NoStop}%
\bibitem [{\citenamefont {Bartok}\ \emph {et~al.}(2010)\citenamefont {Bartok},
  \citenamefont {Payne}, \citenamefont {Kondor},\ and\ \citenamefont
  {Csanyi}}]{bartok2010gaussian}%
  \BibitemOpen
  \bibfield  {author} {\bibinfo {author} {\bibfnamefont {A.~P.}\ \bibnamefont
  {Bartok}}, \bibinfo {author} {\bibfnamefont {M.~C.}\ \bibnamefont {Payne}},
  \bibinfo {author} {\bibfnamefont {R.}~\bibnamefont {Kondor}}, \ and\ \bibinfo
  {author} {\bibfnamefont {G.}~\bibnamefont {Csanyi}},\ }\href@noop {}
  {\bibfield  {journal} {\bibinfo  {journal} {Phys. Rev. Lett.}\ }\textbf
  {\bibinfo {volume} {104}},\ \bibinfo {pages} {136403} (\bibinfo {year}
  {2010})}\BibitemShut {NoStop}%
\bibitem [{\citenamefont {Mones}, \citenamefont {Bernstein},\ and\
  \citenamefont {Csanyi}(2016)}]{mones2016exploration}%
  \BibitemOpen
  \bibfield  {author} {\bibinfo {author} {\bibfnamefont {L.}~\bibnamefont
  {Mones}}, \bibinfo {author} {\bibfnamefont {N.}~\bibnamefont {Bernstein}}, \
  and\ \bibinfo {author} {\bibfnamefont {G.}~\bibnamefont {Csanyi}},\
  }\href@noop {} {\bibfield  {journal} {\bibinfo  {journal} {J. Chem. Theory
  Comput.}\ }\textbf {\bibinfo {volume} {12}},\ \bibinfo {pages} {5100}
  (\bibinfo {year} {2016})}\BibitemShut {NoStop}%
\bibitem [{\citenamefont {Gastegger}, \citenamefont {Behler},\ and\
  \citenamefont {Marquetand}(2017)}]{gastegger2017machine}%
  \BibitemOpen
  \bibfield  {author} {\bibinfo {author} {\bibfnamefont {M.}~\bibnamefont
  {Gastegger}}, \bibinfo {author} {\bibfnamefont {J.}~\bibnamefont {Behler}}, \
  and\ \bibinfo {author} {\bibfnamefont {P.}~\bibnamefont {Marquetand}},\
  }\href {\doibase 10.1039/C7SC02267K} {\bibfield  {journal} {\bibinfo
  {journal} {Chem. Sci.}\ }\textbf {\bibinfo {volume} {8}},\ \bibinfo {pages}
  {6924} (\bibinfo {year} {2017})}\BibitemShut {NoStop}%
\bibitem [{\citenamefont {Kobayashi}\ \emph {et~al.}(2017)\citenamefont
  {Kobayashi}, \citenamefont {Giofr{\'e}}, \citenamefont {Junge}, \citenamefont
  {Ceriotti},\ and\ \citenamefont {Curtin}}]{kobayashi2017neural}%
  \BibitemOpen
  \bibfield  {author} {\bibinfo {author} {\bibfnamefont {R.}~\bibnamefont
  {Kobayashi}}, \bibinfo {author} {\bibfnamefont {D.}~\bibnamefont
  {Giofr{\'e}}}, \bibinfo {author} {\bibfnamefont {T.}~\bibnamefont {Junge}},
  \bibinfo {author} {\bibfnamefont {M.}~\bibnamefont {Ceriotti}}, \ and\
  \bibinfo {author} {\bibfnamefont {W.~A.}\ \bibnamefont {Curtin}},\
  }\href@noop {} {\bibfield  {journal} {\bibinfo  {journal} {Phys. Rev.
  Materials}\ }\textbf {\bibinfo {volume} {1}},\ \bibinfo {pages} {053604}
  (\bibinfo {year} {2017})}\BibitemShut {NoStop}%
\bibitem [{\citenamefont {Carpenter}\ \emph {et~al.}(2017)\citenamefont
  {Carpenter}, \citenamefont {Ezra}, \citenamefont {Farantos}, \citenamefont
  {Kramer},\ and\ \citenamefont {Wiggins}}]{carpenter2017empirical}%
  \BibitemOpen
  \bibfield  {author} {\bibinfo {author} {\bibfnamefont {B.~K.}\ \bibnamefont
  {Carpenter}}, \bibinfo {author} {\bibfnamefont {G.~S.}\ \bibnamefont {Ezra}},
  \bibinfo {author} {\bibfnamefont {S.~C.}\ \bibnamefont {Farantos}}, \bibinfo
  {author} {\bibfnamefont {Z.~C.}\ \bibnamefont {Kramer}}, \ and\ \bibinfo
  {author} {\bibfnamefont {S.}~\bibnamefont {Wiggins}},\ }\href@noop {}
  {\bibfield  {journal} {\bibinfo  {journal} {J. Phys. Chem. B}\ } (\bibinfo
  {year} {2017})}\BibitemShut {NoStop}%
\bibitem [{\citenamefont {Kolb}, \citenamefont {Lentz},\ and\ \citenamefont
  {Kolpak}(2017)}]{kolb2017discovering}%
  \BibitemOpen
  \bibfield  {author} {\bibinfo {author} {\bibfnamefont {B.}~\bibnamefont
  {Kolb}}, \bibinfo {author} {\bibfnamefont {L.~C.}\ \bibnamefont {Lentz}}, \
  and\ \bibinfo {author} {\bibfnamefont {A.~M.}\ \bibnamefont {Kolpak}},\
  }\href@noop {} {\bibfield  {journal} {\bibinfo  {journal} {Sci. Rep.}\
  }\textbf {\bibinfo {volume} {7}},\ \bibinfo {pages} {1192} (\bibinfo {year}
  {2017})}\BibitemShut {NoStop}%
\bibitem [{\citenamefont {Kruglov}\ \emph {et~al.}(2017)\citenamefont
  {Kruglov}, \citenamefont {Sergeev}, \citenamefont {Yanilkin},\ and\
  \citenamefont {Oganov}}]{kruglov2017energy}%
  \BibitemOpen
  \bibfield  {author} {\bibinfo {author} {\bibfnamefont {I.}~\bibnamefont
  {Kruglov}}, \bibinfo {author} {\bibfnamefont {O.}~\bibnamefont {Sergeev}},
  \bibinfo {author} {\bibfnamefont {A.}~\bibnamefont {Yanilkin}}, \ and\
  \bibinfo {author} {\bibfnamefont {A.~R.}\ \bibnamefont {Oganov}},\
  }\href@noop {} {\bibfield  {journal} {\bibinfo  {journal} {Sci. Rep.}\
  }\textbf {\bibinfo {volume} {7}},\ \bibinfo {pages} {8512} (\bibinfo {year}
  {2017})}\BibitemShut {NoStop}%
\bibitem [{\citenamefont {Lubbers}, \citenamefont {Smith},\ and\ \citenamefont
  {Barros}(2017)}]{lubbers2017hierarchical}%
  \BibitemOpen
  \bibfield  {author} {\bibinfo {author} {\bibfnamefont {N.}~\bibnamefont
  {Lubbers}}, \bibinfo {author} {\bibfnamefont {J.~S.}\ \bibnamefont {Smith}},
  \ and\ \bibinfo {author} {\bibfnamefont {K.}~\bibnamefont {Barros}},\
  }\href@noop {} {\bibfield  {journal} {\bibinfo  {journal} {arXiv preprint
  arXiv:1710.00017}\ } (\bibinfo {year} {2017})}\BibitemShut {NoStop}%
\bibitem [{\citenamefont {Mills}, \citenamefont {Spanner},\ and\ \citenamefont
  {Tamblyn}(2017)}]{mills2017deep}%
  \BibitemOpen
  \bibfield  {author} {\bibinfo {author} {\bibfnamefont {K.}~\bibnamefont
  {Mills}}, \bibinfo {author} {\bibfnamefont {M.}~\bibnamefont {Spanner}}, \
  and\ \bibinfo {author} {\bibfnamefont {I.}~\bibnamefont {Tamblyn}},\
  }\href@noop {} {\bibfield  {journal} {\bibinfo  {journal} {Phys. Rev. A}\
  }\textbf {\bibinfo {volume} {96}},\ \bibinfo {pages} {042113} (\bibinfo
  {year} {2017})}\BibitemShut {NoStop}%
\bibitem [{\citenamefont {Wu}, \citenamefont {Shen},\ and\ \citenamefont
  {Yang}(2017)}]{wu2017internal}%
  \BibitemOpen
  \bibfield  {author} {\bibinfo {author} {\bibfnamefont {J.}~\bibnamefont
  {Wu}}, \bibinfo {author} {\bibfnamefont {L.}~\bibnamefont {Shen}}, \ and\
  \bibinfo {author} {\bibfnamefont {W.}~\bibnamefont {Yang}},\ }\href@noop {}
  {\bibfield  {journal} {\bibinfo  {journal} {J. Chem. Phys.}\ }\textbf
  {\bibinfo {volume} {147}},\ \bibinfo {pages} {161732} (\bibinfo {year}
  {2017})}\BibitemShut {NoStop}%
\bibitem [{\citenamefont {Khorshidi}\ and\ \citenamefont
  {Peterson}(2016)}]{khorshidi2016amp}%
  \BibitemOpen
  \bibfield  {author} {\bibinfo {author} {\bibfnamefont {A.}~\bibnamefont
  {Khorshidi}}\ and\ \bibinfo {author} {\bibfnamefont {A.~A.}\ \bibnamefont
  {Peterson}},\ }\href@noop {} {\bibfield  {journal} {\bibinfo  {journal}
  {Comput. Phys. Commun.}\ }\textbf {\bibinfo {volume} {207}},\ \bibinfo
  {pages} {310} (\bibinfo {year} {2016})}\BibitemShut {NoStop}%
\bibitem [{\citenamefont {Shao}\ \emph {et~al.}(2016)\citenamefont {Shao},
  \citenamefont {Chen}, \citenamefont {Zhao},\ and\ \citenamefont
  {Zhang}}]{shao2016communication}%
  \BibitemOpen
  \bibfield  {author} {\bibinfo {author} {\bibfnamefont {K.}~\bibnamefont
  {Shao}}, \bibinfo {author} {\bibfnamefont {J.}~\bibnamefont {Chen}}, \bibinfo
  {author} {\bibfnamefont {Z.}~\bibnamefont {Zhao}}, \ and\ \bibinfo {author}
  {\bibfnamefont {D.~H.}\ \bibnamefont {Zhang}},\ }\href@noop {} {\bibfield
  {journal} {\bibinfo  {journal} {J. Chem. Phys.}\ }\textbf {\bibinfo {volume}
  {145}},\ \bibinfo {pages} {071101} (\bibinfo {year} {2016})}\BibitemShut
  {NoStop}%
\bibitem [{\citenamefont {Zhang}\ and\ \citenamefont
  {Zhang}(2014)}]{zhang2014effects}%
  \BibitemOpen
  \bibfield  {author} {\bibinfo {author} {\bibfnamefont {Z.}~\bibnamefont
  {Zhang}}\ and\ \bibinfo {author} {\bibfnamefont {D.~H.}\ \bibnamefont
  {Zhang}},\ }\href@noop {} {\bibfield  {journal} {\bibinfo  {journal} {J.
  Chem. Phys.}\ }\textbf {\bibinfo {volume} {141}},\ \bibinfo {pages} {144309}
  (\bibinfo {year} {2014})}\BibitemShut {NoStop}%
\bibitem [{\citenamefont {Li}\ \emph {et~al.}(2015)\citenamefont {Li},
  \citenamefont {Chen}, \citenamefont {Zhao}, \citenamefont {Xie},
  \citenamefont {Zhang},\ and\ \citenamefont {Guo}}]{li2015permutationally2}%
  \BibitemOpen
  \bibfield  {author} {\bibinfo {author} {\bibfnamefont {J.}~\bibnamefont
  {Li}}, \bibinfo {author} {\bibfnamefont {J.}~\bibnamefont {Chen}}, \bibinfo
  {author} {\bibfnamefont {Z.}~\bibnamefont {Zhao}}, \bibinfo {author}
  {\bibfnamefont {D.}~\bibnamefont {Xie}}, \bibinfo {author} {\bibfnamefont
  {D.~H.}\ \bibnamefont {Zhang}}, \ and\ \bibinfo {author} {\bibfnamefont
  {H.}~\bibnamefont {Guo}},\ }\href@noop {} {\bibfield  {journal} {\bibinfo
  {journal} {J. Chem. Phys.}\ }\textbf {\bibinfo {volume} {142}},\ \bibinfo
  {pages} {204302} (\bibinfo {year} {2015})}\BibitemShut {NoStop}%
\bibitem [{\citenamefont {Medders}\ \emph {et~al.}(2015)\citenamefont
  {Medders}, \citenamefont {G{\"o}tz}, \citenamefont {Morales}, \citenamefont
  {Bajaj},\ and\ \citenamefont {Paesani}}]{medders2015representation}%
  \BibitemOpen
  \bibfield  {author} {\bibinfo {author} {\bibfnamefont {G.~R.}\ \bibnamefont
  {Medders}}, \bibinfo {author} {\bibfnamefont {A.~W.}\ \bibnamefont
  {G{\"o}tz}}, \bibinfo {author} {\bibfnamefont {M.~A.}\ \bibnamefont
  {Morales}}, \bibinfo {author} {\bibfnamefont {P.}~\bibnamefont {Bajaj}}, \
  and\ \bibinfo {author} {\bibfnamefont {F.}~\bibnamefont {Paesani}},\
  }\href@noop {} {\bibfield  {journal} {\bibinfo  {journal} {J. Chem. Phys.}\
  }\textbf {\bibinfo {volume} {143}},\ \bibinfo {pages} {104102} (\bibinfo
  {year} {2015})}\BibitemShut {NoStop}%
\bibitem [{\citenamefont {Medders}, \citenamefont {Babin},\ and\ \citenamefont
  {Paesani}(2013)}]{medders2013critical}%
  \BibitemOpen
  \bibfield  {author} {\bibinfo {author} {\bibfnamefont {G.~R.}\ \bibnamefont
  {Medders}}, \bibinfo {author} {\bibfnamefont {V.}~\bibnamefont {Babin}}, \
  and\ \bibinfo {author} {\bibfnamefont {F.}~\bibnamefont {Paesani}},\
  }\href@noop {} {\bibfield  {journal} {\bibinfo  {journal} {J. Chem. Theory
  Comput.}\ }\textbf {\bibinfo {volume} {9}},\ \bibinfo {pages} {1103}
  (\bibinfo {year} {2013})}\BibitemShut {NoStop}%
\bibitem [{\citenamefont {Reddy}\ \emph {et~al.}(2016)\citenamefont {Reddy},
  \citenamefont {Straight}, \citenamefont {Bajaj}, \citenamefont {Huy~Pham},
  \citenamefont {Riera}, \citenamefont {Moberg}, \citenamefont {Morales},
  \citenamefont {Knight}, \citenamefont {G{\"o}tz},\ and\ \citenamefont
  {Paesani}}]{reddy2016accuracy}%
  \BibitemOpen
  \bibfield  {author} {\bibinfo {author} {\bibfnamefont {S.~K.}\ \bibnamefont
  {Reddy}}, \bibinfo {author} {\bibfnamefont {S.~C.}\ \bibnamefont {Straight}},
  \bibinfo {author} {\bibfnamefont {P.}~\bibnamefont {Bajaj}}, \bibinfo
  {author} {\bibfnamefont {C.}~\bibnamefont {Huy~Pham}}, \bibinfo {author}
  {\bibfnamefont {M.}~\bibnamefont {Riera}}, \bibinfo {author} {\bibfnamefont
  {D.~R.}\ \bibnamefont {Moberg}}, \bibinfo {author} {\bibfnamefont {M.~A.}\
  \bibnamefont {Morales}}, \bibinfo {author} {\bibfnamefont {C.}~\bibnamefont
  {Knight}}, \bibinfo {author} {\bibfnamefont {A.~W.}\ \bibnamefont
  {G{\"o}tz}}, \ and\ \bibinfo {author} {\bibfnamefont {F.}~\bibnamefont
  {Paesani}},\ }\href@noop {} {\bibfield  {journal} {\bibinfo  {journal} {J.
  Chem. Phys.}\ }\textbf {\bibinfo {volume} {145}},\ \bibinfo {pages} {194504}
  (\bibinfo {year} {2016})}\BibitemShut {NoStop}%
\bibitem [{\citenamefont {Riera}\ \emph {et~al.}(2017)\citenamefont {Riera},
  \citenamefont {Mardirossian}, \citenamefont {Bajaj}, \citenamefont
  {G{\"o}tz},\ and\ \citenamefont {Paesani}}]{riera2017toward}%
  \BibitemOpen
  \bibfield  {author} {\bibinfo {author} {\bibfnamefont {M.}~\bibnamefont
  {Riera}}, \bibinfo {author} {\bibfnamefont {N.}~\bibnamefont {Mardirossian}},
  \bibinfo {author} {\bibfnamefont {P.}~\bibnamefont {Bajaj}}, \bibinfo
  {author} {\bibfnamefont {A.~W.}\ \bibnamefont {G{\"o}tz}}, \ and\ \bibinfo
  {author} {\bibfnamefont {F.}~\bibnamefont {Paesani}},\ }\href@noop {}
  {\bibfield  {journal} {\bibinfo  {journal} {J. Chem. Phys.}\ }\textbf
  {\bibinfo {volume} {147}},\ \bibinfo {pages} {161715} (\bibinfo {year}
  {2017})}\BibitemShut {NoStop}%
\bibitem [{\citenamefont {Moberg}\ \emph {et~al.}(2017)\citenamefont {Moberg},
  \citenamefont {Straight}, \citenamefont {Knight},\ and\ \citenamefont
  {Paesani}}]{moberg2017molecular}%
  \BibitemOpen
  \bibfield  {author} {\bibinfo {author} {\bibfnamefont {D.~R.}\ \bibnamefont
  {Moberg}}, \bibinfo {author} {\bibfnamefont {S.~C.}\ \bibnamefont
  {Straight}}, \bibinfo {author} {\bibfnamefont {C.}~\bibnamefont {Knight}}, \
  and\ \bibinfo {author} {\bibfnamefont {F.}~\bibnamefont {Paesani}},\
  }\href@noop {} {\bibfield  {journal} {\bibinfo  {journal} {J. Phys. Chem.
  Lett.}\ } (\bibinfo {year} {2017})}\BibitemShut {NoStop}%
\bibitem [{\citenamefont {Conte}, \citenamefont {Qu},\ and\ \citenamefont
  {Bowman}(2015)}]{conte2015permutationally}%
  \BibitemOpen
  \bibfield  {author} {\bibinfo {author} {\bibfnamefont {R.}~\bibnamefont
  {Conte}}, \bibinfo {author} {\bibfnamefont {C.}~\bibnamefont {Qu}}, \ and\
  \bibinfo {author} {\bibfnamefont {J.~M.}\ \bibnamefont {Bowman}},\
  }\href@noop {} {\bibfield  {journal} {\bibinfo  {journal} {J. Chem. Theory
  Comput.}\ }\textbf {\bibinfo {volume} {11}},\ \bibinfo {pages} {1631}
  (\bibinfo {year} {2015})}\BibitemShut {NoStop}%
\bibitem [{\citenamefont {Manzhos}, \citenamefont {Dawes},\ and\ \citenamefont
  {Carrington}(2015)}]{manzhos2014neural}%
  \BibitemOpen
  \bibfield  {author} {\bibinfo {author} {\bibfnamefont {S.}~\bibnamefont
  {Manzhos}}, \bibinfo {author} {\bibfnamefont {R.}~\bibnamefont {Dawes}}, \
  and\ \bibinfo {author} {\bibfnamefont {T.}~\bibnamefont {Carrington}},\
  }\href {\doibase 10.1002/qua.24795} {\bibfield  {journal} {\bibinfo
  {journal} {Int. J. Quantum. Chem.}\ }\textbf {\bibinfo {volume} {115}},\
  \bibinfo {pages} {1012} (\bibinfo {year} {2015})}\BibitemShut {NoStop}%
\bibitem [{\citenamefont {Manzhos}, \citenamefont {Yamashita},\ and\
  \citenamefont {Jr.}(2009)}]{manzhos2009fitting}%
  \BibitemOpen
  \bibfield  {author} {\bibinfo {author} {\bibfnamefont {S.}~\bibnamefont
  {Manzhos}}, \bibinfo {author} {\bibfnamefont {K.}~\bibnamefont {Yamashita}},
  \ and\ \bibinfo {author} {\bibfnamefont {T.~C.}\ \bibnamefont {Jr.}},\ }\href
  {\doibase http://dx.doi.org/10.1016/j.cpc.2009.05.022} {\bibfield  {journal}
  {\bibinfo  {journal} {Comput. Phys. Commun.}\ }\textbf {\bibinfo {volume}
  {180}},\ \bibinfo {pages} {2002 } (\bibinfo {year} {2009})}\BibitemShut
  {NoStop}%
\bibitem [{\citenamefont {Malshe}\ \emph {et~al.}(2010)\citenamefont {Malshe},
  \citenamefont {Raff}, \citenamefont {Hagan}, \citenamefont {Bukkapatnam},\
  and\ \citenamefont {Komanduri}}]{malshe2010input}%
  \BibitemOpen
  \bibfield  {author} {\bibinfo {author} {\bibfnamefont {M.}~\bibnamefont
  {Malshe}}, \bibinfo {author} {\bibfnamefont {L.}~\bibnamefont {Raff}},
  \bibinfo {author} {\bibfnamefont {M.}~\bibnamefont {Hagan}}, \bibinfo
  {author} {\bibfnamefont {S.}~\bibnamefont {Bukkapatnam}}, \ and\ \bibinfo
  {author} {\bibfnamefont {R.}~\bibnamefont {Komanduri}},\ }\href@noop {}
  {\bibfield  {journal} {\bibinfo  {journal} {J. Chem. Phys.}\ }\textbf
  {\bibinfo {volume} {132}},\ \bibinfo {pages} {204103} (\bibinfo {year}
  {2010})}\BibitemShut {NoStop}%
\bibitem [{\citenamefont {Peterson}(2016)}]{peterson2016acceleration}%
  \BibitemOpen
  \bibfield  {author} {\bibinfo {author} {\bibfnamefont {A.~A.}\ \bibnamefont
  {Peterson}},\ }\href@noop {} {\bibfield  {journal} {\bibinfo  {journal} {J.
  Chem. Phys.}\ }\textbf {\bibinfo {volume} {145}},\ \bibinfo {pages} {074106}
  (\bibinfo {year} {2016})}\BibitemShut {NoStop}%
\bibitem [{\citenamefont {Piquemal}\ and\ \citenamefont
  {Jordan}(2017)}]{piquemal2017preface}%
  \BibitemOpen
  \bibfield  {author} {\bibinfo {author} {\bibfnamefont {J.-P.}\ \bibnamefont
  {Piquemal}}\ and\ \bibinfo {author} {\bibfnamefont {K.~D.}\ \bibnamefont
  {Jordan}},\ }\href@noop {} {\enquote {\bibinfo {title} {Preface: Special
  topic: From quantum mechanics to force fields},}\ } (\bibinfo {year}
  {2017})\BibitemShut {NoStop}%
\bibitem [{\citenamefont {Cubuk}\ \emph {et~al.}(2017)\citenamefont {Cubuk},
  \citenamefont {Malone}, \citenamefont {Onat}, \citenamefont {Waterland},\
  and\ \citenamefont {Kaxiras}}]{cubuk2017representations}%
  \BibitemOpen
  \bibfield  {author} {\bibinfo {author} {\bibfnamefont {E.~D.}\ \bibnamefont
  {Cubuk}}, \bibinfo {author} {\bibfnamefont {B.~D.}\ \bibnamefont {Malone}},
  \bibinfo {author} {\bibfnamefont {B.}~\bibnamefont {Onat}}, \bibinfo {author}
  {\bibfnamefont {A.}~\bibnamefont {Waterland}}, \ and\ \bibinfo {author}
  {\bibfnamefont {E.}~\bibnamefont {Kaxiras}},\ }\href@noop {} {\bibfield
  {journal} {\bibinfo  {journal} {J. Chem. Phys.}\ }\textbf {\bibinfo {volume}
  {147}},\ \bibinfo {pages} {024104} (\bibinfo {year} {2017})}\BibitemShut
  {NoStop}%
\bibitem [{\citenamefont {John}\ and\ \citenamefont
  {Csanyi}(2017)}]{MBCoarseCsanyi}%
  \BibitemOpen
  \bibfield  {author} {\bibinfo {author} {\bibfnamefont {S.~T.}\ \bibnamefont
  {John}}\ and\ \bibinfo {author} {\bibfnamefont {G.}~\bibnamefont {Csanyi}},\
  }\href {\doibase 10.1021/acs.jpcb.7b09636} {\bibfield  {journal} {\bibinfo
  {journal} {J. Phys. Chem. B}\ } (\bibinfo {year} {2017}),\
  10.1021/acs.jpcb.7b09636},\ \bibinfo {note} {pMID: 29117675},\ \Eprint
  {http://arxiv.org/abs/http://dx.doi.org/10.1021/acs.jpcb.7b09636}
  {http://dx.doi.org/10.1021/acs.jpcb.7b09636} \BibitemShut {NoStop}%
\bibitem [{\citenamefont {Fracchia}\ \emph {et~al.}(2017)\citenamefont
  {Fracchia}, \citenamefont {Del~Frate}, \citenamefont {Mancini}, \citenamefont
  {Rocchia},\ and\ \citenamefont {Barone}}]{FFparafit}%
  \BibitemOpen
  \bibfield  {author} {\bibinfo {author} {\bibfnamefont {F.}~\bibnamefont
  {Fracchia}}, \bibinfo {author} {\bibfnamefont {G.}~\bibnamefont {Del~Frate}},
  \bibinfo {author} {\bibfnamefont {G.}~\bibnamefont {Mancini}}, \bibinfo
  {author} {\bibfnamefont {W.}~\bibnamefont {Rocchia}}, \ and\ \bibinfo
  {author} {\bibfnamefont {V.}~\bibnamefont {Barone}},\ }\href {\doibase
  10.1021/acs.jctc.7b00779} {\bibfield  {journal} {\bibinfo  {journal} {J.
  Chem. Theory Comput.}\ } (\bibinfo {year} {2017}),\
  10.1021/acs.jctc.7b00779},\ \bibinfo {note} {pMID: 29112432},\ \Eprint
  {http://arxiv.org/abs/http://dx.doi.org/10.1021/acs.jctc.7b00779}
  {http://dx.doi.org/10.1021/acs.jctc.7b00779} \BibitemShut {NoStop}%
\bibitem [{\citenamefont {Li}\ \emph {et~al.}(2017)\citenamefont {Li},
  \citenamefont {Li}, \citenamefont {Pickard~IV}, \citenamefont {Narayanan},
  \citenamefont {Sen}, \citenamefont {Chan}, \citenamefont {Sankaranarayanan},
  \citenamefont {Brooks},\ and\ \citenamefont {Roux}}]{li2017machine}%
  \BibitemOpen
  \bibfield  {author} {\bibinfo {author} {\bibfnamefont {Y.}~\bibnamefont
  {Li}}, \bibinfo {author} {\bibfnamefont {H.}~\bibnamefont {Li}}, \bibinfo
  {author} {\bibfnamefont {F.~C.}\ \bibnamefont {Pickard~IV}}, \bibinfo
  {author} {\bibfnamefont {B.}~\bibnamefont {Narayanan}}, \bibinfo {author}
  {\bibfnamefont {F.~G.}\ \bibnamefont {Sen}}, \bibinfo {author} {\bibfnamefont
  {M.~K.}\ \bibnamefont {Chan}}, \bibinfo {author} {\bibfnamefont {S.~K.}\
  \bibnamefont {Sankaranarayanan}}, \bibinfo {author} {\bibfnamefont {B.~R.}\
  \bibnamefont {Brooks}}, \ and\ \bibinfo {author} {\bibfnamefont
  {B.}~\bibnamefont {Roux}},\ }\href@noop {} {\bibfield  {journal} {\bibinfo
  {journal} {J. Chem. Theory Comput.}\ }\textbf {\bibinfo {volume} {13}},\
  \bibinfo {pages} {4492} (\bibinfo {year} {2017})}\BibitemShut {NoStop}%
\bibitem [{\citenamefont {Rupp}\ \emph {et~al.}(2012)\citenamefont {Rupp},
  \citenamefont {Tkatchenko}, \citenamefont {M\"u{}ller},\ and\ \citenamefont
  {Von~Lilienfeld}}]{CM}%
  \BibitemOpen
  \bibfield  {author} {\bibinfo {author} {\bibfnamefont {M.}~\bibnamefont
  {Rupp}}, \bibinfo {author} {\bibfnamefont {A.}~\bibnamefont {Tkatchenko}},
  \bibinfo {author} {\bibfnamefont {K.-R.}\ \bibnamefont {M\"u{}ller}}, \ and\
  \bibinfo {author} {\bibfnamefont {O.~A.}\ \bibnamefont {Von~Lilienfeld}},\
  }\href@noop {} {\bibfield  {journal} {\bibinfo  {journal} {Phys. Rev. Lett.}\
  }\textbf {\bibinfo {volume} {108}},\ \bibinfo {pages} {058301} (\bibinfo
  {year} {2012})}\BibitemShut {NoStop}%
\bibitem [{\citenamefont {Hansen}\ \emph {et~al.}(2015)\citenamefont {Hansen},
  \citenamefont {Biegler}, \citenamefont {Ramakrishnan}, \citenamefont
  {Pronobis}, \citenamefont {Von~Lilienfeld}, \citenamefont {M\"u{}ller},\ and\
  \citenamefont {Tkatchenko}}]{BoBs}%
  \BibitemOpen
  \bibfield  {author} {\bibinfo {author} {\bibfnamefont {K.}~\bibnamefont
  {Hansen}}, \bibinfo {author} {\bibfnamefont {F.}~\bibnamefont {Biegler}},
  \bibinfo {author} {\bibfnamefont {R.}~\bibnamefont {Ramakrishnan}}, \bibinfo
  {author} {\bibfnamefont {W.}~\bibnamefont {Pronobis}}, \bibinfo {author}
  {\bibfnamefont {O.~A.}\ \bibnamefont {Von~Lilienfeld}}, \bibinfo {author}
  {\bibfnamefont {K.-R.}\ \bibnamefont {M\"u{}ller}}, \ and\ \bibinfo {author}
  {\bibfnamefont {A.}~\bibnamefont {Tkatchenko}},\ }\href@noop {} {\bibfield
  {journal} {\bibinfo  {journal} {J. Phys. Chem. Lett.}\ }\textbf {\bibinfo
  {volume} {6}},\ \bibinfo {pages} {2326} (\bibinfo {year} {2015})}\BibitemShut
  {NoStop}%
\bibitem [{\citenamefont {Lopez-Bezanilla}\ and\ \citenamefont {von
  Lilienfeld}(2014)}]{lopez2014modeling}%
  \BibitemOpen
  \bibfield  {author} {\bibinfo {author} {\bibfnamefont {A.}~\bibnamefont
  {Lopez-Bezanilla}}\ and\ \bibinfo {author} {\bibfnamefont {O.~A.}\
  \bibnamefont {von Lilienfeld}},\ }\href@noop {} {\bibfield  {journal}
  {\bibinfo  {journal} {Phys. Rev. B}\ }\textbf {\bibinfo {volume} {89}},\
  \bibinfo {pages} {235411} (\bibinfo {year} {2014})}\BibitemShut {NoStop}%
\bibitem [{\citenamefont {Pilania}\ \emph {et~al.}(2013)\citenamefont
  {Pilania}, \citenamefont {Wang}, \citenamefont {Jiang}, \citenamefont
  {Rajasekaran},\ and\ \citenamefont {Ramprasad}}]{pilania2013accelerating}%
  \BibitemOpen
  \bibfield  {author} {\bibinfo {author} {\bibfnamefont {G.}~\bibnamefont
  {Pilania}}, \bibinfo {author} {\bibfnamefont {C.}~\bibnamefont {Wang}},
  \bibinfo {author} {\bibfnamefont {X.}~\bibnamefont {Jiang}}, \bibinfo
  {author} {\bibfnamefont {S.}~\bibnamefont {Rajasekaran}}, \ and\ \bibinfo
  {author} {\bibfnamefont {R.}~\bibnamefont {Ramprasad}},\ }\href@noop {}
  {\bibfield  {journal} {\bibinfo  {journal} {Sci. Rep.}\ }\textbf {\bibinfo
  {volume} {3}} (\bibinfo {year} {2013})}\BibitemShut {NoStop}%
\bibitem [{\citenamefont {Sch{\"u}tt}\ \emph {et~al.}(2014)\citenamefont
  {Sch{\"u}tt}, \citenamefont {Glawe}, \citenamefont {Brockherde},
  \citenamefont {Sanna}, \citenamefont {M\"u{}ller},\ and\ \citenamefont
  {Gross}}]{schutt2014represent}%
  \BibitemOpen
  \bibfield  {author} {\bibinfo {author} {\bibfnamefont {K.}~\bibnamefont
  {Sch{\"u}tt}}, \bibinfo {author} {\bibfnamefont {H.}~\bibnamefont {Glawe}},
  \bibinfo {author} {\bibfnamefont {F.}~\bibnamefont {Brockherde}}, \bibinfo
  {author} {\bibfnamefont {A.}~\bibnamefont {Sanna}}, \bibinfo {author}
  {\bibfnamefont {K.}~\bibnamefont {M\"u{}ller}}, \ and\ \bibinfo {author}
  {\bibfnamefont {E.}~\bibnamefont {Gross}},\ }\href@noop {} {\bibfield
  {journal} {\bibinfo  {journal} {Phys. Rev. B}\ }\textbf {\bibinfo {volume}
  {89}},\ \bibinfo {pages} {205118} (\bibinfo {year} {2014})}\BibitemShut
  {NoStop}%
\bibitem [{\citenamefont {Ma}\ \emph {et~al.}(2015)\citenamefont {Ma},
  \citenamefont {Li}, \citenamefont {Achenie},\ and\ \citenamefont
  {Xin}}]{ma2015machine}%
  \BibitemOpen
  \bibfield  {author} {\bibinfo {author} {\bibfnamefont {X.}~\bibnamefont
  {Ma}}, \bibinfo {author} {\bibfnamefont {Z.}~\bibnamefont {Li}}, \bibinfo
  {author} {\bibfnamefont {L.~E.}\ \bibnamefont {Achenie}}, \ and\ \bibinfo
  {author} {\bibfnamefont {H.}~\bibnamefont {Xin}},\ }\href@noop {} {\bibfield
  {journal} {\bibinfo  {journal} {J. Phys. Chem. Lett.}\ }\textbf {\bibinfo
  {volume} {6}},\ \bibinfo {pages} {3528} (\bibinfo {year} {2015})}\BibitemShut
  {NoStop}%
\bibitem [{\citenamefont {Nelson}\ \emph {et~al.}(2012)\citenamefont {Nelson},
  \citenamefont {Fernandez-Alberti}, \citenamefont {Chernyak}, \citenamefont
  {Roitberg},\ and\ \citenamefont {Tretiak}}]{nelson2012nonadiabatic}%
  \BibitemOpen
  \bibfield  {author} {\bibinfo {author} {\bibfnamefont {T.}~\bibnamefont
  {Nelson}}, \bibinfo {author} {\bibfnamefont {S.}~\bibnamefont
  {Fernandez-Alberti}}, \bibinfo {author} {\bibfnamefont {V.}~\bibnamefont
  {Chernyak}}, \bibinfo {author} {\bibfnamefont {A.~E.}\ \bibnamefont
  {Roitberg}}, \ and\ \bibinfo {author} {\bibfnamefont {S.}~\bibnamefont
  {Tretiak}},\ }\href@noop {} {\bibfield  {journal} {\bibinfo  {journal} {J.
  Chem. Phys.}\ }\textbf {\bibinfo {volume} {136}},\ \bibinfo {pages} {054108}
  (\bibinfo {year} {2012})}\BibitemShut {NoStop}%
\bibitem [{\citenamefont {Janet}\ and\ \citenamefont
  {Kulik}(2017{\natexlab{a}})}]{janet2017predicting}%
  \BibitemOpen
  \bibfield  {author} {\bibinfo {author} {\bibfnamefont {J.~P.}\ \bibnamefont
  {Janet}}\ and\ \bibinfo {author} {\bibfnamefont {H.~J.}\ \bibnamefont
  {Kulik}},\ }\href@noop {} {\bibfield  {journal} {\bibinfo  {journal} {Chem.
  Sci.}\ } (\bibinfo {year} {2017}{\natexlab{a}})}\BibitemShut {NoStop}%
\bibitem [{\citenamefont {Janet}\ and\ \citenamefont
  {Kulik}(2017{\natexlab{b}})}]{janet2017resolving}%
  \BibitemOpen
  \bibfield  {author} {\bibinfo {author} {\bibfnamefont {J.~P.}\ \bibnamefont
  {Janet}}\ and\ \bibinfo {author} {\bibfnamefont {H.~J.}\ \bibnamefont
  {Kulik}},\ }\href@noop {} {\bibfield  {journal} {\bibinfo  {journal} {J.
  Phys. Chem. A}\ } (\bibinfo {year} {2017}{\natexlab{b}})}\BibitemShut
  {NoStop}%
\bibitem [{\citenamefont {H{\"a}se}, \citenamefont {Kreisbeck},\ and\
  \citenamefont {Aspuru-Guzik}(2017)}]{hase2017machine}%
  \BibitemOpen
  \bibfield  {author} {\bibinfo {author} {\bibfnamefont {F.}~\bibnamefont
  {H{\"a}se}}, \bibinfo {author} {\bibfnamefont {C.}~\bibnamefont {Kreisbeck}},
  \ and\ \bibinfo {author} {\bibfnamefont {A.}~\bibnamefont {Aspuru-Guzik}},\
  }\href@noop {} {\bibfield  {journal} {\bibinfo  {journal} {Chem. Sci.}\ }
  (\bibinfo {year} {2017})}\BibitemShut {NoStop}%
\bibitem [{\citenamefont {McGibbon}\ \emph {et~al.}(2017)\citenamefont
  {McGibbon}, \citenamefont {Taube}, \citenamefont {Donchev}, \citenamefont
  {Siva}, \citenamefont {Hern{\'a}ndez}, \citenamefont {Hargus}, \citenamefont
  {Law}, \citenamefont {Klepeis},\ and\ \citenamefont
  {Shaw}}]{mcgibbon2017improving}%
  \BibitemOpen
  \bibfield  {author} {\bibinfo {author} {\bibfnamefont {R.~T.}\ \bibnamefont
  {McGibbon}}, \bibinfo {author} {\bibfnamefont {A.~G.}\ \bibnamefont {Taube}},
  \bibinfo {author} {\bibfnamefont {A.~G.}\ \bibnamefont {Donchev}}, \bibinfo
  {author} {\bibfnamefont {K.}~\bibnamefont {Siva}}, \bibinfo {author}
  {\bibfnamefont {F.}~\bibnamefont {Hern{\'a}ndez}}, \bibinfo {author}
  {\bibfnamefont {C.}~\bibnamefont {Hargus}}, \bibinfo {author} {\bibfnamefont
  {K.-H.}\ \bibnamefont {Law}}, \bibinfo {author} {\bibfnamefont {J.~L.}\
  \bibnamefont {Klepeis}}, \ and\ \bibinfo {author} {\bibfnamefont {D.~E.}\
  \bibnamefont {Shaw}},\ }\href@noop {} {\bibfield  {journal} {\bibinfo
  {journal} {J. Chem. Phys.}\ }\textbf {\bibinfo {volume} {147}},\ \bibinfo
  {pages} {161725} (\bibinfo {year} {2017})}\BibitemShut {NoStop}%
\bibitem [{\citenamefont {Bereau}\ \emph {et~al.}(2017)\citenamefont {Bereau},
  \citenamefont {DiStasio~Jr}, \citenamefont {Tkatchenko},\ and\ \citenamefont
  {von Lilienfeld}}]{bereau2017non}%
  \BibitemOpen
  \bibfield  {author} {\bibinfo {author} {\bibfnamefont {T.}~\bibnamefont
  {Bereau}}, \bibinfo {author} {\bibfnamefont {R.~A.}\ \bibnamefont
  {DiStasio~Jr}}, \bibinfo {author} {\bibfnamefont {A.}~\bibnamefont
  {Tkatchenko}}, \ and\ \bibinfo {author} {\bibfnamefont {O.~A.}\ \bibnamefont
  {von Lilienfeld}},\ }\href@noop {} {\bibfield  {journal} {\bibinfo  {journal}
  {arXiv preprint arXiv:1710.05871}\ } (\bibinfo {year} {2017})}\BibitemShut
  {NoStop}%
\bibitem [{\citenamefont {Grisafi}\ \emph {et~al.}(2017)\citenamefont
  {Grisafi}, \citenamefont {Wilkins}, \citenamefont {Csanyi},\ and\
  \citenamefont {Ceriotti}}]{grisafi2017symmetry}%
  \BibitemOpen
  \bibfield  {author} {\bibinfo {author} {\bibfnamefont {A.}~\bibnamefont
  {Grisafi}}, \bibinfo {author} {\bibfnamefont {D.~M.}\ \bibnamefont
  {Wilkins}}, \bibinfo {author} {\bibfnamefont {G.}~\bibnamefont {Csanyi}}, \
  and\ \bibinfo {author} {\bibfnamefont {M.}~\bibnamefont {Ceriotti}},\
  }\href@noop {} {\bibfield  {journal} {\bibinfo  {journal} {arXiv preprint
  arXiv:1709.06757}\ } (\bibinfo {year} {2017})}\BibitemShut {NoStop}%
\bibitem [{\citenamefont {Isayev}\ \emph {et~al.}(2017)\citenamefont {Isayev},
  \citenamefont {Oses}, \citenamefont {Toher}, \citenamefont {Gossett},
  \citenamefont {Curtarolo},\ and\ \citenamefont
  {Tropsha}}]{isayev2017universal}%
  \BibitemOpen
  \bibfield  {author} {\bibinfo {author} {\bibfnamefont {O.}~\bibnamefont
  {Isayev}}, \bibinfo {author} {\bibfnamefont {C.}~\bibnamefont {Oses}},
  \bibinfo {author} {\bibfnamefont {C.}~\bibnamefont {Toher}}, \bibinfo
  {author} {\bibfnamefont {E.}~\bibnamefont {Gossett}}, \bibinfo {author}
  {\bibfnamefont {S.}~\bibnamefont {Curtarolo}}, \ and\ \bibinfo {author}
  {\bibfnamefont {A.}~\bibnamefont {Tropsha}},\ }\href@noop {} {\bibfield
  {journal} {\bibinfo  {journal} {Nat. Commun.}\ }\textbf {\bibinfo {volume}
  {8}},\ \bibinfo {pages} {15679} (\bibinfo {year} {2017})}\BibitemShut
  {NoStop}%
\bibitem [{\citenamefont {Ghiringhelli}\ \emph {et~al.}(2017)\citenamefont
  {Ghiringhelli}, \citenamefont {Vybiral}, \citenamefont {Ahmetcik},
  \citenamefont {Ouyang}, \citenamefont {Levchenko}, \citenamefont {Draxl},\
  and\ \citenamefont {Scheffler}}]{ghiringhelli2017learning}%
  \BibitemOpen
  \bibfield  {author} {\bibinfo {author} {\bibfnamefont {L.~M.}\ \bibnamefont
  {Ghiringhelli}}, \bibinfo {author} {\bibfnamefont {J.}~\bibnamefont
  {Vybiral}}, \bibinfo {author} {\bibfnamefont {E.}~\bibnamefont {Ahmetcik}},
  \bibinfo {author} {\bibfnamefont {R.}~\bibnamefont {Ouyang}}, \bibinfo
  {author} {\bibfnamefont {S.~V.}\ \bibnamefont {Levchenko}}, \bibinfo {author}
  {\bibfnamefont {C.}~\bibnamefont {Draxl}}, \ and\ \bibinfo {author}
  {\bibfnamefont {M.}~\bibnamefont {Scheffler}},\ }\href@noop {} {\bibfield
  {journal} {\bibinfo  {journal} {New J. Phys.}\ }\textbf {\bibinfo {volume}
  {19}},\ \bibinfo {pages} {023017} (\bibinfo {year} {2017})}\BibitemShut
  {NoStop}%
\bibitem [{\citenamefont {Ouyang}\ \emph {et~al.}(2017)\citenamefont {Ouyang},
  \citenamefont {Curtarolo}, \citenamefont {Ahmetcik}, \citenamefont
  {Scheffler},\ and\ \citenamefont {Ghiringhelli}}]{ouyang2017sisso}%
  \BibitemOpen
  \bibfield  {author} {\bibinfo {author} {\bibfnamefont {R.}~\bibnamefont
  {Ouyang}}, \bibinfo {author} {\bibfnamefont {S.}~\bibnamefont {Curtarolo}},
  \bibinfo {author} {\bibfnamefont {E.}~\bibnamefont {Ahmetcik}}, \bibinfo
  {author} {\bibfnamefont {M.}~\bibnamefont {Scheffler}}, \ and\ \bibinfo
  {author} {\bibfnamefont {L.~M.}\ \bibnamefont {Ghiringhelli}},\ }\href@noop
  {} {\bibfield  {journal} {\bibinfo  {journal} {arXiv preprint
  arXiv:1710.03319}\ } (\bibinfo {year} {2017})}\BibitemShut {NoStop}%
\bibitem [{\citenamefont {Faber}\ \emph {et~al.}(2017)\citenamefont {Faber},
  \citenamefont {Hutchison}, \citenamefont {Huang}, \citenamefont {Gilmer},
  \citenamefont {Schoenholz}, \citenamefont {Dahl}, \citenamefont {Vinyals},
  \citenamefont {Kearnes}, \citenamefont {Riley},\ and\ \citenamefont {von
  Lilienfeld}}]{faber2017prediction}%
  \BibitemOpen
  \bibfield  {author} {\bibinfo {author} {\bibfnamefont {F.~A.}\ \bibnamefont
  {Faber}}, \bibinfo {author} {\bibfnamefont {L.}~\bibnamefont {Hutchison}},
  \bibinfo {author} {\bibfnamefont {B.}~\bibnamefont {Huang}}, \bibinfo
  {author} {\bibfnamefont {J.}~\bibnamefont {Gilmer}}, \bibinfo {author}
  {\bibfnamefont {S.~S.}\ \bibnamefont {Schoenholz}}, \bibinfo {author}
  {\bibfnamefont {G.~E.}\ \bibnamefont {Dahl}}, \bibinfo {author}
  {\bibfnamefont {O.}~\bibnamefont {Vinyals}}, \bibinfo {author} {\bibfnamefont
  {S.}~\bibnamefont {Kearnes}}, \bibinfo {author} {\bibfnamefont {P.~F.}\
  \bibnamefont {Riley}}, \ and\ \bibinfo {author} {\bibfnamefont {O.~A.}\
  \bibnamefont {von Lilienfeld}},\ }\href@noop {} {\bibfield  {journal}
  {\bibinfo  {journal} {J. Chem. Theory Comput.}\ } (\bibinfo {year}
  {2017})}\BibitemShut {NoStop}%
\bibitem [{\citenamefont {Sch{\"u}tt}\ \emph {et~al.}(2017)\citenamefont
  {Sch{\"u}tt}, \citenamefont {Arbabzadah}, \citenamefont {Chmiela},
  \citenamefont {M{\"u}ller},\ and\ \citenamefont
  {Tkatchenko}}]{schutt2017quantum}%
  \BibitemOpen
  \bibfield  {author} {\bibinfo {author} {\bibfnamefont {K.~T.}\ \bibnamefont
  {Sch{\"u}tt}}, \bibinfo {author} {\bibfnamefont {F.}~\bibnamefont
  {Arbabzadah}}, \bibinfo {author} {\bibfnamefont {S.}~\bibnamefont {Chmiela}},
  \bibinfo {author} {\bibfnamefont {K.~R.}\ \bibnamefont {M{\"u}ller}}, \ and\
  \bibinfo {author} {\bibfnamefont {A.}~\bibnamefont {Tkatchenko}},\
  }\href@noop {} {\bibfield  {journal} {\bibinfo  {journal} {Nat. Commun.}\
  }\textbf {\bibinfo {volume} {8}},\ \bibinfo {pages} {13890} (\bibinfo {year}
  {2017})}\BibitemShut {NoStop}%
\bibitem [{\citenamefont {Zhou}\ \emph {et~al.}(2017)\citenamefont {Zhou},
  \citenamefont {Zeng}, \citenamefont {Chi}, \citenamefont {Luo}, \citenamefont
  {Liu}, \citenamefont {Zhan}, \citenamefont {He},\ and\ \citenamefont
  {Zhang}}]{MSLSTM}%
  \BibitemOpen
  \bibfield  {author} {\bibinfo {author} {\bibfnamefont {X.-X.}\ \bibnamefont
  {Zhou}}, \bibinfo {author} {\bibfnamefont {W.-F.}\ \bibnamefont {Zeng}},
  \bibinfo {author} {\bibfnamefont {H.}~\bibnamefont {Chi}}, \bibinfo {author}
  {\bibfnamefont {C.}~\bibnamefont {Luo}}, \bibinfo {author} {\bibfnamefont
  {C.}~\bibnamefont {Liu}}, \bibinfo {author} {\bibfnamefont {J.}~\bibnamefont
  {Zhan}}, \bibinfo {author} {\bibfnamefont {S.-M.}\ \bibnamefont {He}}, \ and\
  \bibinfo {author} {\bibfnamefont {Z.}~\bibnamefont {Zhang}},\ }\href
  {\doibase 10.1021/acs.analchem.7b02566} {\bibfield  {journal} {\bibinfo
  {journal} {Anal. Chem.}\ } (\bibinfo {year} {2017}),\
  10.1021/acs.analchem.7b02566},\ \bibinfo {note} {pMID: 29125736},\ \Eprint
  {http://arxiv.org/abs/http://dx.doi.org/10.1021/acs.analchem.7b02566}
  {http://dx.doi.org/10.1021/acs.analchem.7b02566} \BibitemShut {NoStop}%
\bibitem [{\citenamefont {Timoshenko}\ \emph {et~al.}(2017)\citenamefont
  {Timoshenko}, \citenamefont {Lu}, \citenamefont {Lin},\ and\ \citenamefont
  {Frenkel}}]{timoshenko2017supervised}%
  \BibitemOpen
  \bibfield  {author} {\bibinfo {author} {\bibfnamefont {J.}~\bibnamefont
  {Timoshenko}}, \bibinfo {author} {\bibfnamefont {D.}~\bibnamefont {Lu}},
  \bibinfo {author} {\bibfnamefont {Y.}~\bibnamefont {Lin}}, \ and\ \bibinfo
  {author} {\bibfnamefont {A.~I.}\ \bibnamefont {Frenkel}},\ }\href@noop {}
  {\bibfield  {journal} {\bibinfo  {journal} {J. Phys. Chem. Lett.}\ }
  (\bibinfo {year} {2017})}\BibitemShut {NoStop}%
\bibitem [{\citenamefont {Li}, \citenamefont {Cai},\ and\ \citenamefont
  {He}(2017)}]{li2017learning}%
  \BibitemOpen
  \bibfield  {author} {\bibinfo {author} {\bibfnamefont {J.}~\bibnamefont
  {Li}}, \bibinfo {author} {\bibfnamefont {D.}~\bibnamefont {Cai}}, \ and\
  \bibinfo {author} {\bibfnamefont {X.}~\bibnamefont {He}},\ }\href@noop {}
  {\bibfield  {journal} {\bibinfo  {journal} {arXiv preprint arXiv:1709.03741}\
  } (\bibinfo {year} {2017})}\BibitemShut {NoStop}%
\bibitem [{\citenamefont {Ramsundar}\ \emph {et~al.}(2017)\citenamefont
  {Ramsundar}, \citenamefont {Liu}, \citenamefont {Wu}, \citenamefont {Verras},
  \citenamefont {Tudor}, \citenamefont {Sheridan},\ and\ \citenamefont
  {Pande}}]{ramsundar2017multitask}%
  \BibitemOpen
  \bibfield  {author} {\bibinfo {author} {\bibfnamefont {B.}~\bibnamefont
  {Ramsundar}}, \bibinfo {author} {\bibfnamefont {B.}~\bibnamefont {Liu}},
  \bibinfo {author} {\bibfnamefont {Z.}~\bibnamefont {Wu}}, \bibinfo {author}
  {\bibfnamefont {A.}~\bibnamefont {Verras}}, \bibinfo {author} {\bibfnamefont
  {M.}~\bibnamefont {Tudor}}, \bibinfo {author} {\bibfnamefont {R.~P.}\
  \bibnamefont {Sheridan}}, \ and\ \bibinfo {author} {\bibfnamefont
  {V.}~\bibnamefont {Pande}},\ }\href@noop {} {\bibfield  {journal} {\bibinfo
  {journal} {J. Chem. Inf. Model.}\ }\textbf {\bibinfo {volume} {57}},\
  \bibinfo {pages} {2068} (\bibinfo {year} {2017})}\BibitemShut {NoStop}%
\bibitem [{\citenamefont {Hachmann}\ \emph {et~al.}(2011)\citenamefont
  {Hachmann}, \citenamefont {Olivares-Amaya}, \citenamefont {Atahan-Evrenk},
  \citenamefont {Amador-Bedolla}, \citenamefont {S{\'a}nchez-Carrera},
  \citenamefont {Gold-Parker}, \citenamefont {Vogt}, \citenamefont {Brockway},\
  and\ \citenamefont {Aspuru-Guzik}}]{hachmann2011harvard}%
  \BibitemOpen
  \bibfield  {author} {\bibinfo {author} {\bibfnamefont {J.}~\bibnamefont
  {Hachmann}}, \bibinfo {author} {\bibfnamefont {R.}~\bibnamefont
  {Olivares-Amaya}}, \bibinfo {author} {\bibfnamefont {S.}~\bibnamefont
  {Atahan-Evrenk}}, \bibinfo {author} {\bibfnamefont {C.}~\bibnamefont
  {Amador-Bedolla}}, \bibinfo {author} {\bibfnamefont {R.~S.}\ \bibnamefont
  {S{\'a}nchez-Carrera}}, \bibinfo {author} {\bibfnamefont {A.}~\bibnamefont
  {Gold-Parker}}, \bibinfo {author} {\bibfnamefont {L.}~\bibnamefont {Vogt}},
  \bibinfo {author} {\bibfnamefont {A.~M.}\ \bibnamefont {Brockway}}, \ and\
  \bibinfo {author} {\bibfnamefont {A.}~\bibnamefont {Aspuru-Guzik}},\
  }\href@noop {} {\bibfield  {journal} {\bibinfo  {journal} {J. Phys. Chem.
  Lett.}\ }\textbf {\bibinfo {volume} {2}},\ \bibinfo {pages} {2241} (\bibinfo
  {year} {2011})}\BibitemShut {NoStop}%
\bibitem [{\citenamefont {Hachmann}\ \emph {et~al.}(2014)\citenamefont
  {Hachmann}, \citenamefont {Olivares-Amaya}, \citenamefont {Jinich},
  \citenamefont {Appleton}, \citenamefont {Blood-Forsythe}, \citenamefont
  {Seress}, \citenamefont {Roman-Salgado}, \citenamefont {Trepte},
  \citenamefont {Atahan-Evrenk},\ and\ \citenamefont {Er}}]{hachmann2014lead}%
  \BibitemOpen
  \bibfield  {author} {\bibinfo {author} {\bibfnamefont {J.}~\bibnamefont
  {Hachmann}}, \bibinfo {author} {\bibfnamefont {R.}~\bibnamefont
  {Olivares-Amaya}}, \bibinfo {author} {\bibfnamefont {A.}~\bibnamefont
  {Jinich}}, \bibinfo {author} {\bibfnamefont {A.~L.}\ \bibnamefont
  {Appleton}}, \bibinfo {author} {\bibfnamefont {M.~A.}\ \bibnamefont
  {Blood-Forsythe}}, \bibinfo {author} {\bibfnamefont {L.~R.}\ \bibnamefont
  {Seress}}, \bibinfo {author} {\bibfnamefont {C.}~\bibnamefont
  {Roman-Salgado}}, \bibinfo {author} {\bibfnamefont {K.}~\bibnamefont
  {Trepte}}, \bibinfo {author} {\bibfnamefont {S.}~\bibnamefont
  {Atahan-Evrenk}}, \ and\ \bibinfo {author} {\bibfnamefont {S.}~\bibnamefont
  {Er}},\ }\href@noop {} {\bibfield  {journal} {\bibinfo  {journal} {Energ.
  Environ. Sci.}\ }\textbf {\bibinfo {volume} {7}},\ \bibinfo {pages} {698}
  (\bibinfo {year} {2014})}\BibitemShut {NoStop}%
\bibitem [{\citenamefont {Isayev}\ \emph {et~al.}(2015)\citenamefont {Isayev},
  \citenamefont {Fourches}, \citenamefont {Muratov}, \citenamefont {Oses},
  \citenamefont {Rasch}, \citenamefont {Tropsha},\ and\ \citenamefont
  {Curtarolo}}]{isayev2015materials}%
  \BibitemOpen
  \bibfield  {author} {\bibinfo {author} {\bibfnamefont {O.}~\bibnamefont
  {Isayev}}, \bibinfo {author} {\bibfnamefont {D.}~\bibnamefont {Fourches}},
  \bibinfo {author} {\bibfnamefont {E.~N.}\ \bibnamefont {Muratov}}, \bibinfo
  {author} {\bibfnamefont {C.}~\bibnamefont {Oses}}, \bibinfo {author}
  {\bibfnamefont {K.}~\bibnamefont {Rasch}}, \bibinfo {author} {\bibfnamefont
  {A.}~\bibnamefont {Tropsha}}, \ and\ \bibinfo {author} {\bibfnamefont
  {S.}~\bibnamefont {Curtarolo}},\ }\href {\doibase 10.1021/cm503507h}
  {\bibfield  {journal} {\bibinfo  {journal} {Chem. Mater.}\ }\textbf {\bibinfo
  {volume} {27}},\ \bibinfo {pages} {735} (\bibinfo {year} {2015})},\ \Eprint
  {http://arxiv.org/abs/http://dx.doi.org/10.1021/cm503507h}
  {http://dx.doi.org/10.1021/cm503507h} \BibitemShut {NoStop}%
\bibitem [{\citenamefont {Kim}\ \emph {et~al.}(2017)\citenamefont {Kim},
  \citenamefont {Huang}, \citenamefont {Tomala}, \citenamefont {Matthews},
  \citenamefont {Strubell}, \citenamefont {Saunders}, \citenamefont
  {McCallum},\ and\ \citenamefont {Olivetti}}]{kim2017machine}%
  \BibitemOpen
  \bibfield  {author} {\bibinfo {author} {\bibfnamefont {E.}~\bibnamefont
  {Kim}}, \bibinfo {author} {\bibfnamefont {K.}~\bibnamefont {Huang}}, \bibinfo
  {author} {\bibfnamefont {A.}~\bibnamefont {Tomala}}, \bibinfo {author}
  {\bibfnamefont {S.}~\bibnamefont {Matthews}}, \bibinfo {author}
  {\bibfnamefont {E.}~\bibnamefont {Strubell}}, \bibinfo {author}
  {\bibfnamefont {A.}~\bibnamefont {Saunders}}, \bibinfo {author}
  {\bibfnamefont {A.}~\bibnamefont {McCallum}}, \ and\ \bibinfo {author}
  {\bibfnamefont {E.}~\bibnamefont {Olivetti}},\ }\href@noop {} {\bibfield
  {journal} {\bibinfo  {journal} {Sci. Data}\ }\textbf {\bibinfo {volume}
  {4}},\ \bibinfo {pages} {sdata2017127} (\bibinfo {year} {2017})}\BibitemShut
  {NoStop}%
\bibitem [{\citenamefont {Segler}, \citenamefont {Preu{\ss}},\ and\
  \citenamefont {Waller}(2017)}]{segler2017towards}%
  \BibitemOpen
  \bibfield  {author} {\bibinfo {author} {\bibfnamefont {M.}~\bibnamefont
  {Segler}}, \bibinfo {author} {\bibfnamefont {M.}~\bibnamefont {Preu{\ss}}}, \
  and\ \bibinfo {author} {\bibfnamefont {M.~P.}\ \bibnamefont {Waller}},\
  }\href@noop {} {\bibfield  {journal} {\bibinfo  {journal} {arXiv preprint
  arXiv:1702.00020}\ } (\bibinfo {year} {2017})}\BibitemShut {NoStop}%
\bibitem [{\citenamefont {Olivares-Amaya}\ \emph {et~al.}(2011)\citenamefont
  {Olivares-Amaya}, \citenamefont {Amador-Bedolla}, \citenamefont {Hachmann},
  \citenamefont {Atahan-Evrenk}, \citenamefont {S{\'a}nchez-Carrera},
  \citenamefont {Vogt},\ and\ \citenamefont
  {Aspuru-Guzik}}]{olivares2011accelerated}%
  \BibitemOpen
  \bibfield  {author} {\bibinfo {author} {\bibfnamefont {R.}~\bibnamefont
  {Olivares-Amaya}}, \bibinfo {author} {\bibfnamefont {C.}~\bibnamefont
  {Amador-Bedolla}}, \bibinfo {author} {\bibfnamefont {J.}~\bibnamefont
  {Hachmann}}, \bibinfo {author} {\bibfnamefont {S.}~\bibnamefont
  {Atahan-Evrenk}}, \bibinfo {author} {\bibfnamefont {R.~S.}\ \bibnamefont
  {S{\'a}nchez-Carrera}}, \bibinfo {author} {\bibfnamefont {L.}~\bibnamefont
  {Vogt}}, \ and\ \bibinfo {author} {\bibfnamefont {A.}~\bibnamefont
  {Aspuru-Guzik}},\ }\href@noop {} {\bibfield  {journal} {\bibinfo  {journal}
  {Energ. Environ. Sci.}\ }\textbf {\bibinfo {volume} {4}},\ \bibinfo {pages}
  {4849} (\bibinfo {year} {2011})}\BibitemShut {NoStop}%
\bibitem [{\citenamefont {Guimaraes}\ \emph {et~al.}(2017)\citenamefont
  {Guimaraes}, \citenamefont {Sanchez-Lengeling}, \citenamefont {Farias},\ and\
  \citenamefont {Aspuru-Guzik}}]{guimaraes2017objective}%
  \BibitemOpen
  \bibfield  {author} {\bibinfo {author} {\bibfnamefont {G.~L.}\ \bibnamefont
  {Guimaraes}}, \bibinfo {author} {\bibfnamefont {B.}~\bibnamefont
  {Sanchez-Lengeling}}, \bibinfo {author} {\bibfnamefont {P.~L.~C.}\
  \bibnamefont {Farias}}, \ and\ \bibinfo {author} {\bibfnamefont
  {A.}~\bibnamefont {Aspuru-Guzik}},\ }\href@noop {} {\bibfield  {journal}
  {\bibinfo  {journal} {arXiv preprint arXiv:1705.10843}\ } (\bibinfo {year}
  {2017})}\BibitemShut {NoStop}%
\bibitem [{\citenamefont {Wei}, \citenamefont {Duvenaud},\ and\ \citenamefont
  {Aspuru-Guzik}(2016)}]{wei2016neural}%
  \BibitemOpen
  \bibfield  {author} {\bibinfo {author} {\bibfnamefont {J.~N.}\ \bibnamefont
  {Wei}}, \bibinfo {author} {\bibfnamefont {D.}~\bibnamefont {Duvenaud}}, \
  and\ \bibinfo {author} {\bibfnamefont {A.}~\bibnamefont {Aspuru-Guzik}},\
  }\href@noop {} {\bibfield  {journal} {\bibinfo  {journal} {ACS Cent. Sci.}\
  }\textbf {\bibinfo {volume} {2}},\ \bibinfo {pages} {725} (\bibinfo {year}
  {2016})}\BibitemShut {NoStop}%
\bibitem [{\citenamefont {G{\'o}mez-Bombarelli}\ \emph
  {et~al.}(2016)\citenamefont {G{\'o}mez-Bombarelli}, \citenamefont {Duvenaud},
  \citenamefont {Hern{\'a}ndez-Lobato}, \citenamefont {Aguilera-Iparraguirre},
  \citenamefont {Hirzel}, \citenamefont {Adams},\ and\ \citenamefont
  {Aspuru-Guzik}}]{gomez2016automatic}%
  \BibitemOpen
  \bibfield  {author} {\bibinfo {author} {\bibfnamefont {R.}~\bibnamefont
  {G{\'o}mez-Bombarelli}}, \bibinfo {author} {\bibfnamefont {D.}~\bibnamefont
  {Duvenaud}}, \bibinfo {author} {\bibfnamefont {J.~M.}\ \bibnamefont
  {Hern{\'a}ndez-Lobato}}, \bibinfo {author} {\bibfnamefont {J.}~\bibnamefont
  {Aguilera-Iparraguirre}}, \bibinfo {author} {\bibfnamefont {T.~D.}\
  \bibnamefont {Hirzel}}, \bibinfo {author} {\bibfnamefont {R.~P.}\
  \bibnamefont {Adams}}, \ and\ \bibinfo {author} {\bibfnamefont
  {A.}~\bibnamefont {Aspuru-Guzik}},\ }\href@noop {} {\bibfield  {journal}
  {\bibinfo  {journal} {arXiv preprint arXiv:1610.02415}\ } (\bibinfo {year}
  {2016})}\BibitemShut {NoStop}%
\bibitem [{\citenamefont {Jinnouchi}\ and\ \citenamefont
  {Asahi}(2017)}]{jinnouchi2017predicting}%
  \BibitemOpen
  \bibfield  {author} {\bibinfo {author} {\bibfnamefont {R.}~\bibnamefont
  {Jinnouchi}}\ and\ \bibinfo {author} {\bibfnamefont {R.}~\bibnamefont
  {Asahi}},\ }\href@noop {} {\bibfield  {journal} {\bibinfo  {journal} {J.
  Phys. Chem. Lett.}\ }\textbf {\bibinfo {volume} {8}},\ \bibinfo {pages}
  {4279} (\bibinfo {year} {2017})}\BibitemShut {NoStop}%
\bibitem [{\citenamefont {Ulissi}\ \emph {et~al.}(2017)\citenamefont {Ulissi},
  \citenamefont {Tang}, \citenamefont {Xiao}, \citenamefont {Liu},
  \citenamefont {Torelli}, \citenamefont {Karamad}, \citenamefont {Cummins},
  \citenamefont {Hahn}, \citenamefont {Lewis}, \citenamefont {Jaramillo} \emph
  {et~al.}}]{ulissi2017machine}%
  \BibitemOpen
  \bibfield  {author} {\bibinfo {author} {\bibfnamefont {Z.~W.}\ \bibnamefont
  {Ulissi}}, \bibinfo {author} {\bibfnamefont {M.~T.}\ \bibnamefont {Tang}},
  \bibinfo {author} {\bibfnamefont {J.}~\bibnamefont {Xiao}}, \bibinfo {author}
  {\bibfnamefont {X.}~\bibnamefont {Liu}}, \bibinfo {author} {\bibfnamefont
  {D.~A.}\ \bibnamefont {Torelli}}, \bibinfo {author} {\bibfnamefont
  {M.}~\bibnamefont {Karamad}}, \bibinfo {author} {\bibfnamefont
  {K.}~\bibnamefont {Cummins}}, \bibinfo {author} {\bibfnamefont
  {C.}~\bibnamefont {Hahn}}, \bibinfo {author} {\bibfnamefont {N.~S.}\
  \bibnamefont {Lewis}}, \bibinfo {author} {\bibfnamefont {T.~F.}\ \bibnamefont
  {Jaramillo}},  \emph {et~al.},\ }\href@noop {} {\bibfield  {journal}
  {\bibinfo  {journal} {ACS Catal.}\ }\textbf {\bibinfo {volume} {7}},\
  \bibinfo {pages} {6600} (\bibinfo {year} {2017})}\BibitemShut {NoStop}%
\bibitem [{\citenamefont {Sun}\ \emph {et~al.}(2017)\citenamefont {Sun},
  \citenamefont {Bai}, \citenamefont {Li},\ and\ \citenamefont
  {Wang}}]{sun2017machine}%
  \BibitemOpen
  \bibfield  {author} {\bibinfo {author} {\bibfnamefont {Y.~T.}\ \bibnamefont
  {Sun}}, \bibinfo {author} {\bibfnamefont {H.}~\bibnamefont {Bai}}, \bibinfo
  {author} {\bibfnamefont {M.-Z.}\ \bibnamefont {Li}}, \ and\ \bibinfo {author}
  {\bibfnamefont {W.}~\bibnamefont {Wang}},\ }\href@noop {} {\bibfield
  {journal} {\bibinfo  {journal} {J. Phys. Chem. Lett.}\ }\textbf {\bibinfo
  {volume} {8}},\ \bibinfo {pages} {3434} (\bibinfo {year} {2017})}\BibitemShut
  {NoStop}%
\bibitem [{\citenamefont {Grimme}(2006)}]{grimme2006semiempirical}%
  \BibitemOpen
  \bibfield  {author} {\bibinfo {author} {\bibfnamefont {S.}~\bibnamefont
  {Grimme}},\ }\href@noop {} {\bibfield  {journal} {\bibinfo  {journal} {J.
  Comput. Chem.}\ }\textbf {\bibinfo {volume} {27}},\ \bibinfo {pages} {1787}
  (\bibinfo {year} {2006})}\BibitemShut {NoStop}%
\bibitem [{\citenamefont {Thole}(1981)}]{thole1981molecular}%
  \BibitemOpen
  \bibfield  {author} {\bibinfo {author} {\bibfnamefont {B.~T.}\ \bibnamefont
  {Thole}},\ }\href@noop {} {\bibfield  {journal} {\bibinfo  {journal} {Chem.
  Phys.}\ }\textbf {\bibinfo {volume} {59}},\ \bibinfo {pages} {341} (\bibinfo
  {year} {1981})}\BibitemShut {NoStop}%
\bibitem [{\citenamefont {Smith}, \citenamefont {Isayev},\ and\ \citenamefont
  {Roitberg}(2017)}]{smith2017ani}%
  \BibitemOpen
  \bibfield  {author} {\bibinfo {author} {\bibfnamefont {J.~S.}\ \bibnamefont
  {Smith}}, \bibinfo {author} {\bibfnamefont {O.}~\bibnamefont {Isayev}}, \
  and\ \bibinfo {author} {\bibfnamefont {A.~E.}\ \bibnamefont {Roitberg}},\
  }\href@noop {} {\bibfield  {journal} {\bibinfo  {journal} {Chem. Sci.}\ }
  (\bibinfo {year} {2017})}\BibitemShut {NoStop}%
\bibitem [{\citenamefont {Chai}\ and\ \citenamefont
  {Head-Gordon}(2008)}]{chai2008long}%
  \BibitemOpen
  \bibfield  {author} {\bibinfo {author} {\bibfnamefont {J.-D.}\ \bibnamefont
  {Chai}}\ and\ \bibinfo {author} {\bibfnamefont {M.}~\bibnamefont
  {Head-Gordon}},\ }\href@noop {} {\bibfield  {journal} {\bibinfo  {journal}
  {Phys. Chem. Chem. Phys.}\ }\textbf {\bibinfo {volume} {10}},\ \bibinfo
  {pages} {6615} (\bibinfo {year} {2008})}\BibitemShut {NoStop}%
\bibitem [{\citenamefont {Ceriotti}, \citenamefont {More},\ and\ \citenamefont
  {Manolopoulos}(2014)}]{ceriotti2014pi}%
  \BibitemOpen
  \bibfield  {author} {\bibinfo {author} {\bibfnamefont {M.}~\bibnamefont
  {Ceriotti}}, \bibinfo {author} {\bibfnamefont {J.}~\bibnamefont {More}}, \
  and\ \bibinfo {author} {\bibfnamefont {D.~E.}\ \bibnamefont {Manolopoulos}},\
  }\href@noop {} {\bibfield  {journal} {\bibinfo  {journal} {Comput. Phys.
  Commun}\ }\textbf {\bibinfo {volume} {185}},\ \bibinfo {pages} {1019}
  (\bibinfo {year} {2014})}\BibitemShut {NoStop}%
\bibitem [{\citenamefont {Deringer}\ and\ \citenamefont
  {Csanyi}(2017)}]{deringer2017machine}%
  \BibitemOpen
  \bibfield  {author} {\bibinfo {author} {\bibfnamefont {V.~L.}\ \bibnamefont
  {Deringer}}\ and\ \bibinfo {author} {\bibfnamefont {G.}~\bibnamefont
  {Csanyi}},\ }\href@noop {} {\bibfield  {journal} {\bibinfo  {journal} {Phys.
  Rev. B}\ }\textbf {\bibinfo {volume} {95}},\ \bibinfo {pages} {094203}
  (\bibinfo {year} {2017})}\BibitemShut {NoStop}%
\bibitem [{\citenamefont {Morawietz}\ and\ \citenamefont
  {Behler}(2013)}]{morawietz2013density}%
  \BibitemOpen
  \bibfield  {author} {\bibinfo {author} {\bibfnamefont {T.}~\bibnamefont
  {Morawietz}}\ and\ \bibinfo {author} {\bibfnamefont {J.}~\bibnamefont
  {Behler}},\ }\href@noop {} {\bibfield  {journal} {\bibinfo  {journal} {J.
  Phys. Chem. A}\ }\textbf {\bibinfo {volume} {117}},\ \bibinfo {pages} {7356}
  (\bibinfo {year} {2013})}\BibitemShut {NoStop}%
\bibitem [{\citenamefont {Abadi}\ \emph {et~al.}(2015)\citenamefont {Abadi},
  \citenamefont {Agarwal}, \citenamefont {Barham}, \citenamefont {Brevdo},
  \citenamefont {Chen}, \citenamefont {Citro}, \citenamefont {Corrado},
  \citenamefont {Davis}, \citenamefont {Dean}, \citenamefont {Devin},
  \citenamefont {Ghemawat}, \citenamefont {Goodfellow}, \citenamefont {Harp},
  \citenamefont {Irving}, \citenamefont {Isard}, \citenamefont {Jia},
  \citenamefont {Jozefowicz}, \citenamefont {Kaiser}, \citenamefont {Kudlur},
  \citenamefont {Levenberg}, \citenamefont {Man\'{e}}, \citenamefont {Monga},
  \citenamefont {Moore}, \citenamefont {Murray}, \citenamefont {Olah},
  \citenamefont {Schuster}, \citenamefont {Shlens}, \citenamefont {Steiner},
  \citenamefont {Sutskever}, \citenamefont {Talwar}, \citenamefont {Tucker},
  \citenamefont {Vanhoucke}, \citenamefont {Vasudevan}, \citenamefont
  {Vi\'{e}gas}, \citenamefont {Vinyals}, \citenamefont {Warden}, \citenamefont
  {Wattenberg}, \citenamefont {Wicke}, \citenamefont {Yu},\ and\ \citenamefont
  {Zheng}}]{tensorflow2015-whitepaper}%
  \BibitemOpen
  \bibfield  {author} {\bibinfo {author} {\bibfnamefont {M.}~\bibnamefont
  {Abadi}}, \bibinfo {author} {\bibfnamefont {A.}~\bibnamefont {Agarwal}},
  \bibinfo {author} {\bibfnamefont {P.}~\bibnamefont {Barham}}, \bibinfo
  {author} {\bibfnamefont {E.}~\bibnamefont {Brevdo}}, \bibinfo {author}
  {\bibfnamefont {Z.}~\bibnamefont {Chen}}, \bibinfo {author} {\bibfnamefont
  {C.}~\bibnamefont {Citro}}, \bibinfo {author} {\bibfnamefont {G.~S.}\
  \bibnamefont {Corrado}}, \bibinfo {author} {\bibfnamefont {A.}~\bibnamefont
  {Davis}}, \bibinfo {author} {\bibfnamefont {J.}~\bibnamefont {Dean}},
  \bibinfo {author} {\bibfnamefont {M.}~\bibnamefont {Devin}}, \bibinfo
  {author} {\bibfnamefont {S.}~\bibnamefont {Ghemawat}}, \bibinfo {author}
  {\bibfnamefont {I.}~\bibnamefont {Goodfellow}}, \bibinfo {author}
  {\bibfnamefont {A.}~\bibnamefont {Harp}}, \bibinfo {author} {\bibfnamefont
  {G.}~\bibnamefont {Irving}}, \bibinfo {author} {\bibfnamefont
  {M.}~\bibnamefont {Isard}}, \bibinfo {author} {\bibfnamefont
  {Y.}~\bibnamefont {Jia}}, \bibinfo {author} {\bibfnamefont {R.}~\bibnamefont
  {Jozefowicz}}, \bibinfo {author} {\bibfnamefont {L.}~\bibnamefont {Kaiser}},
  \bibinfo {author} {\bibfnamefont {M.}~\bibnamefont {Kudlur}}, \bibinfo
  {author} {\bibfnamefont {J.}~\bibnamefont {Levenberg}}, \bibinfo {author}
  {\bibfnamefont {D.}~\bibnamefont {Man\'{e}}}, \bibinfo {author}
  {\bibfnamefont {R.}~\bibnamefont {Monga}}, \bibinfo {author} {\bibfnamefont
  {S.}~\bibnamefont {Moore}}, \bibinfo {author} {\bibfnamefont
  {D.}~\bibnamefont {Murray}}, \bibinfo {author} {\bibfnamefont
  {C.}~\bibnamefont {Olah}}, \bibinfo {author} {\bibfnamefont {M.}~\bibnamefont
  {Schuster}}, \bibinfo {author} {\bibfnamefont {J.}~\bibnamefont {Shlens}},
  \bibinfo {author} {\bibfnamefont {B.}~\bibnamefont {Steiner}}, \bibinfo
  {author} {\bibfnamefont {I.}~\bibnamefont {Sutskever}}, \bibinfo {author}
  {\bibfnamefont {K.}~\bibnamefont {Talwar}}, \bibinfo {author} {\bibfnamefont
  {P.}~\bibnamefont {Tucker}}, \bibinfo {author} {\bibfnamefont
  {V.}~\bibnamefont {Vanhoucke}}, \bibinfo {author} {\bibfnamefont
  {V.}~\bibnamefont {Vasudevan}}, \bibinfo {author} {\bibfnamefont
  {F.}~\bibnamefont {Vi\'{e}gas}}, \bibinfo {author} {\bibfnamefont
  {O.}~\bibnamefont {Vinyals}}, \bibinfo {author} {\bibfnamefont
  {P.}~\bibnamefont {Warden}}, \bibinfo {author} {\bibfnamefont
  {M.}~\bibnamefont {Wattenberg}}, \bibinfo {author} {\bibfnamefont
  {M.}~\bibnamefont {Wicke}}, \bibinfo {author} {\bibfnamefont
  {Y.}~\bibnamefont {Yu}}, \ and\ \bibinfo {author} {\bibfnamefont
  {X.}~\bibnamefont {Zheng}},\ }\href {http://tensorflow.org/} {\enquote
  {\bibinfo {title} {{TensorFlow}: Large-scale machine learning on
  heterogeneous systems},}\ } (\bibinfo {year} {2015}),\ \bibinfo {note}
  {software available from tensorflow.org}\BibitemShut {NoStop}%
\bibitem [{\citenamefont {Fennell}\ and\ \citenamefont
  {Gezelter}(2006)}]{fennell2006ewald}%
  \BibitemOpen
  \bibfield  {author} {\bibinfo {author} {\bibfnamefont {C.~J.}\ \bibnamefont
  {Fennell}}\ and\ \bibinfo {author} {\bibfnamefont {J.~D.}\ \bibnamefont
  {Gezelter}},\ }\href@noop {} {\bibfield  {journal} {\bibinfo  {journal} {J.
  Chem. Phys.}\ }\textbf {\bibinfo {volume} {124}},\ \bibinfo {pages} {234104}
  (\bibinfo {year} {2006})}\BibitemShut {NoStop}%
\bibitem [{\citenamefont {Shao}\ \emph {et~al.}(2015)\citenamefont {Shao},
  \citenamefont {Gan}, \citenamefont {Epifanovsky}, \citenamefont {Gilbert},
  \citenamefont {Wormit}, \citenamefont {Kussmann}, \citenamefont {Lange},
  \citenamefont {Behn}, \citenamefont {Deng}, \citenamefont {Feng} \emph
  {et~al.}}]{shao2015advances}%
  \BibitemOpen
  \bibfield  {author} {\bibinfo {author} {\bibfnamefont {Y.}~\bibnamefont
  {Shao}}, \bibinfo {author} {\bibfnamefont {Z.}~\bibnamefont {Gan}}, \bibinfo
  {author} {\bibfnamefont {E.}~\bibnamefont {Epifanovsky}}, \bibinfo {author}
  {\bibfnamefont {A.~T.}\ \bibnamefont {Gilbert}}, \bibinfo {author}
  {\bibfnamefont {M.}~\bibnamefont {Wormit}}, \bibinfo {author} {\bibfnamefont
  {J.}~\bibnamefont {Kussmann}}, \bibinfo {author} {\bibfnamefont {A.~W.}\
  \bibnamefont {Lange}}, \bibinfo {author} {\bibfnamefont {A.}~\bibnamefont
  {Behn}}, \bibinfo {author} {\bibfnamefont {J.}~\bibnamefont {Deng}}, \bibinfo
  {author} {\bibfnamefont {X.}~\bibnamefont {Feng}},  \emph {et~al.},\
  }\href@noop {} {\bibfield  {journal} {\bibinfo  {journal} {Mol. Phys.}\
  }\textbf {\bibinfo {volume} {113}},\ \bibinfo {pages} {184} (\bibinfo {year}
  {2015})}\BibitemShut {NoStop}%
\bibitem [{\citenamefont {Srivastava}\ \emph {et~al.}(2014)\citenamefont
  {Srivastava}, \citenamefont {Hinton}, \citenamefont {Krizhevsky},
  \citenamefont {Sutskever},\ and\ \citenamefont
  {Salakhutdinov}}]{srivastava2014dropout}%
  \BibitemOpen
  \bibfield  {author} {\bibinfo {author} {\bibfnamefont {N.}~\bibnamefont
  {Srivastava}}, \bibinfo {author} {\bibfnamefont {G.~E.}\ \bibnamefont
  {Hinton}}, \bibinfo {author} {\bibfnamefont {A.}~\bibnamefont {Krizhevsky}},
  \bibinfo {author} {\bibfnamefont {I.}~\bibnamefont {Sutskever}}, \ and\
  \bibinfo {author} {\bibfnamefont {R.}~\bibnamefont {Salakhutdinov}},\
  }\href@noop {} {\bibfield  {journal} {\bibinfo  {journal} {J. Mach. Learn.
  Res.}\ }\textbf {\bibinfo {volume} {15}},\ \bibinfo {pages} {1929} (\bibinfo
  {year} {2014})}\BibitemShut {NoStop}%
\bibitem [{\citenamefont {Kingma}\ and\ \citenamefont
  {Ba}(2014)}]{kingma2014adam}%
  \BibitemOpen
  \bibfield  {author} {\bibinfo {author} {\bibfnamefont {D.}~\bibnamefont
  {Kingma}}\ and\ \bibinfo {author} {\bibfnamefont {J.}~\bibnamefont {Ba}},\
  }\href@noop {} {\bibfield  {journal} {\bibinfo  {journal} {arXiv preprint
  arXiv:1412.6980}\ } (\bibinfo {year} {2014})}\BibitemShut {NoStop}%
\bibitem [{\citenamefont {Henkelman}, \citenamefont {Uberuaga},\ and\
  \citenamefont {J{\'o}nsson}(2000)}]{henkelman2000climbing}%
  \BibitemOpen
  \bibfield  {author} {\bibinfo {author} {\bibfnamefont {G.}~\bibnamefont
  {Henkelman}}, \bibinfo {author} {\bibfnamefont {B.~P.}\ \bibnamefont
  {Uberuaga}}, \ and\ \bibinfo {author} {\bibfnamefont {H.}~\bibnamefont
  {J{\'o}nsson}},\ }\href@noop {} {\bibfield  {journal} {\bibinfo  {journal}
  {J. Chem. Phys.}\ }\textbf {\bibinfo {volume} {113}},\ \bibinfo {pages}
  {9901} (\bibinfo {year} {2000})}\BibitemShut {NoStop}%
\bibitem [{\citenamefont {Barducci}, \citenamefont {Bussi},\ and\ \citenamefont
  {Parrinello}(2008)}]{barducci2008well}%
  \BibitemOpen
  \bibfield  {author} {\bibinfo {author} {\bibfnamefont {A.}~\bibnamefont
  {Barducci}}, \bibinfo {author} {\bibfnamefont {G.}~\bibnamefont {Bussi}}, \
  and\ \bibinfo {author} {\bibfnamefont {M.}~\bibnamefont {Parrinello}},\
  }\href@noop {} {\bibfield  {journal} {\bibinfo  {journal} {Phys. Rev. Lett.}\
  }\textbf {\bibinfo {volume} {100}},\ \bibinfo {pages} {020603} (\bibinfo
  {year} {2008})}\BibitemShut {NoStop}%
\bibitem [{\citenamefont {Tkatchenko}\ \emph {et~al.}(2012)\citenamefont
  {Tkatchenko}, \citenamefont {DiStasio~Jr}, \citenamefont {Car},\ and\
  \citenamefont {Scheffler}}]{tkatchenko2012accurate}%
  \BibitemOpen
  \bibfield  {author} {\bibinfo {author} {\bibfnamefont {A.}~\bibnamefont
  {Tkatchenko}}, \bibinfo {author} {\bibfnamefont {R.~A.}\ \bibnamefont
  {DiStasio~Jr}}, \bibinfo {author} {\bibfnamefont {R.}~\bibnamefont {Car}}, \
  and\ \bibinfo {author} {\bibfnamefont {M.}~\bibnamefont {Scheffler}},\
  }\href@noop {} {\bibfield  {journal} {\bibinfo  {journal} {Phys. Rev. Lett.}\
  }\textbf {\bibinfo {volume} {108}},\ \bibinfo {pages} {236402} (\bibinfo
  {year} {2012})}\BibitemShut {NoStop}%
\bibitem [{\citenamefont {Eastman}\ \emph {et~al.}(2017)\citenamefont
  {Eastman}, \citenamefont {Swails}, \citenamefont {Chodera}, \citenamefont
  {McGibbon}, \citenamefont {Zhao}, \citenamefont {Beauchamp}, \citenamefont
  {Wang}, \citenamefont {Simmonett}, \citenamefont {Harrigan}, \citenamefont
  {Stern}, \citenamefont {Wiewiora}, \citenamefont {Brooks},\ and\
  \citenamefont {Pande}}]{OpenMM}%
  \BibitemOpen
  \bibfield  {author} {\bibinfo {author} {\bibfnamefont {P.}~\bibnamefont
  {Eastman}}, \bibinfo {author} {\bibfnamefont {J.}~\bibnamefont {Swails}},
  \bibinfo {author} {\bibfnamefont {J.~D.}\ \bibnamefont {Chodera}}, \bibinfo
  {author} {\bibfnamefont {R.~T.}\ \bibnamefont {McGibbon}}, \bibinfo {author}
  {\bibfnamefont {Y.}~\bibnamefont {Zhao}}, \bibinfo {author} {\bibfnamefont
  {K.~A.}\ \bibnamefont {Beauchamp}}, \bibinfo {author} {\bibfnamefont {L.-P.}\
  \bibnamefont {Wang}}, \bibinfo {author} {\bibfnamefont {A.~C.}\ \bibnamefont
  {Simmonett}}, \bibinfo {author} {\bibfnamefont {M.~P.}\ \bibnamefont
  {Harrigan}}, \bibinfo {author} {\bibfnamefont {C.~D.}\ \bibnamefont {Stern}},
  \bibinfo {author} {\bibfnamefont {R.~P.}\ \bibnamefont {Wiewiora}}, \bibinfo
  {author} {\bibfnamefont {B.~R.}\ \bibnamefont {Brooks}}, \ and\ \bibinfo
  {author} {\bibfnamefont {V.~S.}\ \bibnamefont {Pande}},\ }\href {\doibase
  10.1371/journal.pcbi.1005659} {\bibfield  {journal} {\bibinfo  {journal}
  {PLOS Computational Biology}\ }\textbf {\bibinfo {volume} {13}},\ \bibinfo
  {pages} {1} (\bibinfo {year} {2017})}\BibitemShut {NoStop}%
\end{thebibliography}

\clearpage

\end{document}